\documentstyle[11pt]{l-aa}
\input psfig
\def\Real{{\rm I\mathchoice{\kern-0.70mm}{\kern-0.70mm}{\kern-0.65mm}%
  {\kern-0.50mm}R}}

\font \bolditalics = cmmib10
\def\bx#1{\leavevmode\thinspace\hbox{\vrule\vtop{\vbox{\hrule\kern1pt
        \hbox{\vphantom{\tt/}\thinspace{\bf#1}\thinspace}}
      \kern1pt\hrule}\vrule}\thinspace}

\def \vc #1{{\textfont1=\bolditalics \hbox{$\bf#1$}}}

\def\thetag{{\vc \theta}}

\begin{document}
\thesaurus{12.03.3 - 12.04.1 - 12.07.1 - 11.03.4 MS 1008-1224}
\title{Weak lensing analysis of MS 1008-1224 with the VLT\thanks{Based on
observations obtained at the Very Large Telescope at
Cerro Paranal operated by the European Southern Observatory.}}
\author{R. Athreya$^{1}$, Y. Mellier$^{1,2}$, L. van Waerbeke$^{3}$,
B. Fort$^{1}$, R. Pell\'o$^{4}$, M. Dantel-Fort$^{2}$.}
\institute{$^{1}$Institut d'Astrophysique de Paris, 98 bis boulevard 
Arago, 75014 Paris, France,\\ 
$^{2}$Observatoire de Paris, DEMIRM, 61 Av. de 
l'Observatoire, 75014 Paris, France,\\
$^{3}$Canadian Institute for Theoretical Astrophysics, 60 St 
George Str., Toronto, M5S 3H8, Canada\\
$^{4}$Observatoire Midi-Pyr\'en\'ees, UMR 5572, 14 avenue
\'Edouard Belin, 31400 Toulouse, France.\\
}
\offprints{(Ramana Athreya) athreya@iap.fr}
\date{Received xx; accepted xx}
\maketitle

\begin{abstract}
We present a gravitational lensing analysis of the cluster of galaxies MS 
1008-1224 ($z=0.30$), based on very deep observations obtained using the VLT 
with FORS and ISAAC during the science verification phase.\\
We reconstructed the projected mass distribution from the B, V, R and I band
data using two different algorithms independently. The mass maps are remarkably 
similar, which confirmed that the PSF correction worked well, thanks to the 
superb quality of the images.\\
The J and K band data (ISAAC) were combined with the BVRI (FORS) data to measure
the photometric redshifts of galaxies inside the ISAAC field and to constrain 
the redshift distribution of the lensed sources. This enabled us to scale the 
gravitational convergence into an accurate mass estimate.\\
The total mass inferred from weak shear is 2.3 $\times$ 10$^{14}$ $h^{-1}$ 
M$_{\odot}$ on large scales (within $\approx 700$ $h^{-1}$ kpc) which is in good
agreement with the X-ray mass. The Mass-to-light ratios are also in excellent 
agreement (319$h$, against 312$h$ from the X-ray). The measured mass profile is 
well fit by both Navarro, Frenk and White and isothermal sphere with core radius
models although the NFW appears to be slightly better.\\
In the inner regions, the lensing mass is about 2 times higher than the X-ray 
mass, which supports the long-held view that complex physical processes occuring
in the innermost parts of lensing-clusters are responsible for the X-ray/lensing
mass discrepancy. We found that the central part of the cluster is composed of 
two mass peaks whose the center of mass is located 15 arcsecond north of the cD 
galaxy. This provides an explanation for the 15 arcsecond offset between the cD 
and the center of the X-ray map already reported elsewhere.\\
The optical, X-ray and the mass distributions show that MS 1008-1224 is composed
of many subsystems which are probably undergoing a merger. It is likely that the
gas is not in equilibrium in the innermost regions which vitiates the X-ray mass
estimate.\\
MS 1008-1224 shows a remarkable case of cluster-cluster lensing. The photometric
redshifts show an excess of galaxies located 30 arcseconds south-west of the
cD galaxy at a redshift of about 0.9. This distant cluster is lensed by MS 
1008-1224, which enables the detection of many of its galaxies. Hence, MS 
1008-1224 can be viewed as a gravitational telescope facilitating the study of a
distant cluster.\\
These results show that detailed investigations of lensing clusters require very
deep imaging with sub-arcsecond seeing in multiple bands (BVRI and JK).
Our analysis demonstrates that a thorough investigation of clusters of galaxies 
and a careful handling of the biases cannot be performed without a dataset which
fulfills these requirements. The outstanding capabilities of the VLT at Cerro 
Paranal make it a unique tool which makes such studies possible.


\end{abstract}

\section{Introduction} 
The analysis of the distribution of dark matter in clusters of galaxies provides
an important insight into the history of structure formation in the Universe. 
The epoch of formation of clusters and their evolution with redshift are 
dependent on cosmological parameters and the nature of (dark) matter. For 
example, the existence of even a few massive clusters at redshift z = 1 would be
a strong pointer to a low-$\Omega$ universes.

In order to follow the cosmic history of clusters with look-back time, it is 
important to have reliable tools at hand to study the nature of matter (the 
mass, its distribution, the fraction of baryonic and non-baryonic matter etc).
Gravitational lensing and bremsstrahlung emission from hot intra-cluster gas are
two processes which help us probe these issues. Unfortunately, the results from
these two approaches have not yet provided a mutually consistent picture.

Indeed, X-ray mass estimates show discrepancies with weak and strong lensing 
mass estimates of clusters of galaxies.  The origin of the discrepancy is not 
yet fully understood (see Mellier 1999 for a review).  The total mass inferred 
from lensing exceeds the X-ray mass by a factor of about two for some clusters 
including well studied clusters like A2218 (Miralda-Escud\'e \& Babul 1995). 
Investigations of a dozen clusters by Smail et al (1997), Allen (1998) or Lewis 
et al (1999) have not illuminated the reasons for the discrepancy. In an attempt
to solve the problem, Allen (1998) compared cooling flow and non-cooling flow 
clusters and observed that the former do not show the mass discrepancy. This 
result suggests that the assumptions regarding the dynamical and thermal state 
of the hot intra-cluster gas, a key ingredient for the X-ray mass estimate, are 
not realistic enough to result in a satisfactory model of non-cooling flow (i.e.
presumably non-relaxed) clusters of galaxies. This interpretation is confirmed 
by B\"ohringer et al (1998) who found an excellent agreement between the X-ray, 
the strong and the weak lensing analyses of the relaxed cluster A2390. However, 
Lewis et al (1999) did not find similar trends in their cluster sample. They 
argue that even some cooling flow clusters show significant discrepancies 
between X-ray and lensing mass, particularly with strong lensing estimates.  

It is likely that this mass discrepancy is the result of several other
less-than-valid assumptions. For example, the comparison between X-ray and weak 
gravitational lensing is done by extrapolating the best fit of the  X-ray 
profile far beyond the region where data are reliable, where uncertainties are 
obviously significant and the shape of the (assumed) analytical profile used for
extrapolation also has a considerable impact on the mass estimate (Lewis et al,
1999;  B\"ohringer et al, 1999).

Lensing mass estimates are not free from bias either. N-body simulations by Cen 
(1997) and Metzler (1999) show that projection effects of in-falling filaments 
of matter towards the cluster centre can significantly bias the projected mass 
density inferred from weak lensing analysis to values higher than from X-ray.  
The amplitude of the bias seems to range between 10 and 20 per cent, but   
more generally, projection effects generated  by any structures along the line 
of sight, can overestimate the total mass by about 30 per cent (Reblinsky \& 
Bartelmann 1999). It is therefore important to explore ways which could improve 
the accuracy of the total mass estimate from lensing inversion algorithms in 
order to disentangle the systematics from random errors and to separate the 
contributions of gravitational lensing analyses to the discrepancy from those of
the assumptions in the X-ray analyses.

Very deep observations of clusters of galaxies in multiple bands and with 
subarcsecond seeing can considerably improve the reliability of mass estimates 
from weak lensing; the depth increases the number density of lensed galaxies
thereby improving the resolution of the mass reconstruction; multicolor
observations allow estimation of photometric redshifts to be obtain the redshift
distribution of background sources; finally, subarcsecond seeing makes for a
good determination of object shapes and accurate PSF correction. Rarely are all
of these stringent requirements satisfied simultaneously in ground based 
observations. Fortunately, the recent observations obtained during the 
Science Verification Programme\footnote{The data were made publicly 
available to the ESO community in May 1999. Details of this dataset may be 
obtained from the URL http://www.hq.eso.org/science/ut1sv}  
for the FORS (FOcal Reducer/low dispersion Spectrograph; Appenzeller et al 1998)
and the ISAAC (Infrared Spectrometer and Array Camera; Moorwood et al 1999) 
instruments mounted 
on the first Very Large Telescope, UT1/ANTU, provide an excellent 
dataset on the lensing cluster MS 1008-1224.  The images obtained at
Paranal have a seeing better than 0.7 arcsecond in B, V, R and I bands 
over a 6 arcminute field of view, as well as on the 2.5 arcminute field 
covering the central part of the cluster
in J and K-bands. The quality and depth of these VLT images are therefore among
the best data ever obtained from the ground  for weak lensing mass 
reconstruction of a cluster.

MS 1008-1224 is a galaxy cluster selected from the Einstein 
Medium Sensitivity Survey (Gioia \& Luppino 1994). 
It is one of the 16 EMSS clusters observed by Le F\`evre et al
(1994) in which they found strong lensing features. The cluster is at 
redshift $z=0.3062$ (Lewis et al, 1999) and is dominated by a cD galaxy.
The X-ray luminosity is $L_X$(0.3-3.5 keV)=4.5 $\times 10^{44}$ erg.s$^{-1}$ 
(From Lewis et al 1999, with $H_0$=100 km s$^{-1}$ Mpc$^{-1}$ and $q_0=0.1$) 
and its temperature inferred from ASCA observation is $T_X=7.29$ keV
(Mushotzky \& Scharf 1997). According to Lewis et al (1999), the X-ray contours
are centered 15 arcseconds to the north of the cD and show a rather circular
pattern outward with an extension towards the north.

The paper is organized as follows :  Section 2 describes the observations, the
data analyses and some optical properties of the VLT images, including the 
description of some lensed features and magnified distant objects. The mass 
reconstructions, from weak shear analysis as well as from depletion produced by 
magnification bias, are presented in Section 3. The results are discussed in 
Section 4. The main conclusions of the study are summarized in Section 5.

In this paper we have used $H_0=100h^{-1}$ kms$^{-1}$ Mpc$^{-1}$, $\Omega_0=0.3$
$\Lambda$ = 0. This corresponds to a typical scale of 176 $h^{-1}$ kpc for 1 
arcminute at the redshift of the cluster.

\section{Optical properties of the lens MS 1008-1224}
The observations were all carried out by the Science Verification Team at ESO. 
The details of the data processing, from the image acquisition to co-addition 
and calibration may be found at the ESO web site (see URL in the previous 
section). Table \ref{vltsummary} is a summary of the most important
characteristics. The 6\farcm{8} $\times$ 6\farcm8 field of view of FORS 
corresponds to a physical size of 1.2$h^{-1}$ Mpc at redshift $z=0.31$. The VLT 
BVRI data provide a global view of the morphology of MS 1008-1224 and of the 
field of galaxies around it. The central region of the FORS field is shown in 
Fig. \ref{isaacfield}. The gravitational distortion pattern is obvious even on a
casual inspection of the VLT images. Together with the X-ray observations of 
Lewis et al  (1999), these high quality multi-band images are well suited for a 
thorough analysis of the distribution of mass within the cluster.

Source detection and photometry were performed in a standard manner
using the latest version of the release 2.1.0 of the SExtractor software
(Bertin \& Arnoults, 1996) which is publicly available on the TERAPIX
web site (http://terapix.iap.fr/sextractor/). The magnitude distributions of
galaxies from this field are shown in Figure \ref{magdist}.
The stars in the field were identified by their location on the
magnitude - half-light radius plot (see Fahlman et al 1994). 

\subsection{Lensed features}
The field shows many lensing features which are good candidates for 
spectroscopy. Some of the more interesting ones are discussed below. The most 
obvious central arc is shown in Fig. \ref{isaacfield} and Fig. \ref{arc4bvrijk}.
Unfortunately, from its morphology alone, it is not easy to identify the lens 
configuration. Its shape and its position very close to the cD suggest that 
this is a radial arc, but its shape on none of the images shows a double 
structure which would convincingly demonstrate that it is composed of two 
merging parts of the same source. Alternatively, it could be a tangential arc. 
In that case, its curvature indicates that the main deflector is to its north. 
This would imply that the cD is not located at the center of the potential well.
Further, an inspection of its image in different bands (Fig. \ref{arc4bvrijk})
shows that we cannot even exclude that this simply comprises galaxies 
superimposed along the line of sight.

Some other systems look like typical arcs, like object \#2 reported by Le 
F\`evre et al (1994) and  shown on Fig. \ref{isaacfield} and Fig. 
\ref{arc2bvrijk}. The magnification reveals many structures inside. The 
irregularities are indications that this is probably a late-type system. Its 
redshift is lower than 4, since it is visible on the B-image.

We now have spectroscopic observations which show that object \#1 in Fig. 
\ref{isaacfield}, identified as an arc by Le F\`evre et al (1994), is actually 
an edge-on spiral within the cluster itself. The $B-\alpha I$ image (Fig. 
\ref{isaacfield}) shows its central part is circular and very red, in contrast 
to its extended and elliptical periphery which is much bluer, as expected for 
the disk of an edge-on spiral.

Fig. \ref{obj13bvrijk} shows  a remarkable system which seems to be a magnified 
dropout candidate. It is located close to a bright cluster galaxy, so it is 
probably magnified twice, by the cluster as well as by its neighbour on the
image. Object \#14, located close to object \#2 (Fig. \ref{arc2bvrijk}) also
seems to be a very high redshift lensed galaxy.

There are many gravitational pairs in the field. The most obvious one is object 
\#5, with two images of apparently reversed parity. However, the colours of the 
two objects do not provide conclusive evidence that they are indeed a pair and
they could well be unrelated objects. Object \#3 also seems to be a 
gravitational pair. However, we have not found a third image for either of 
these pairs.


\begin{figure*}
\begin{center}
\caption{\label{isaacfield} A B-$\alpha$I image of the central parts of the
FORS field showing the cluster centre. The square marks the field covered by
ISAAC. The coefficient $\alpha$ has been optimized to suppress the central cD 
and other bright cluster galaxies. This procedure reveals more clearly the 
very red objects which appear as white features on the image. Conversely, the 
darkest objects are the bluest galaxies in the field.  The arrow at the 
centre points to the location of (the residual of) the cD. The removal of the
bright cluster members makes the background galaxies more visible.
The circles indicate some of the interesting features and are labeled with the
same numbers as in Le F\`evre et al (1994). Object \#1 is the arc candidate 
reported by Le F\`evre et al which our spectroscopic observation shows is an
edge-on galaxy in the cluster. Object 2 is the other arc candidate reported by
Le F\`evre et al. Objects \#3 and \#5 are two pair candidates and object \#4 is 
a radial arc located close to the cD. However, its strong curvature towards the
north suggests that it could well be a tangential arc produced by a 
concentration of 
matter there. Only a spectroscopic follow-up can establish the possible 
gravitational nature of these objects. 
}
\end{center}
\end{figure*}
\begin{figure*}
\begin{center}
\caption{\label{arc4bvrijk}
 Zooms of the BVRIJK images of the central arc \#4. From left to right: B, V, 
R, I, J and K image. Its ambiguous shape makes it difficult to decide if it is 
a radial or a tangential arc. It could be also interacting galaxies or galaxies 
projected along the same line of sight by accident. This is certainly a system 
for which spectroscopic data will be useful.
}
\end{center}
\end{figure*}
\begin{figure*}
\begin{center}
\psfig{figure=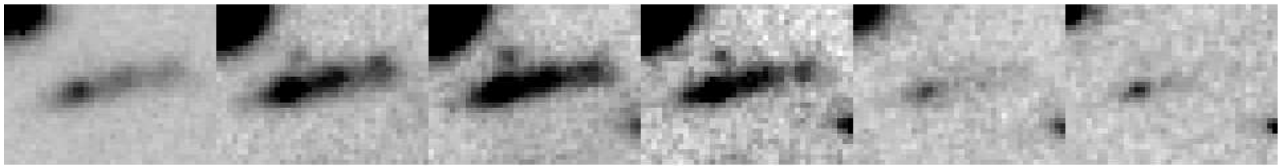,width=17.cm}
\caption{\label{arc2bvrijk}
 Zooms of the BVRIJK images of the arc \#2 reported by Le F\`evre et al (1994).
Thanks to the magnification, many structures are visible inside the arc. Note 
also the small object just above the arc. Note the object \#14 at the 
bottom-right edge of the field. This is likely to be a dropout candidate lensed
by the cluster.
}
\end{center}
\end{figure*}
\begin{figure*}
\begin{center}
\psfig{figure=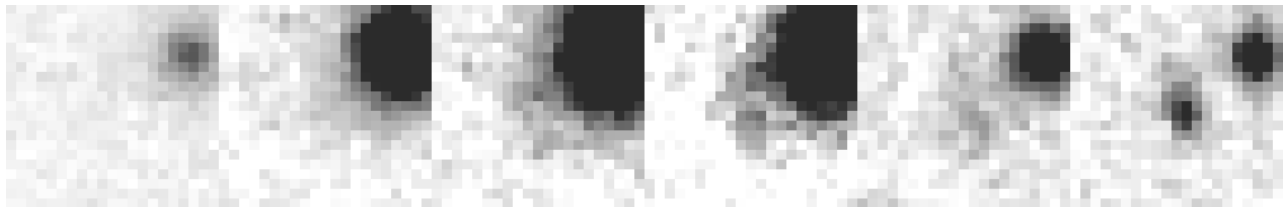,width=17.cm}
\caption{\label{obj13bvrijk}
Zooms of the BVRIJK images of the object \#13. This is a typical dropout system
which is not visible in $B$, so its redshift is probably larger than 4. The 
system is very close to a cluster galaxy and has probably been magnified twice,
by the cluster and by this galaxy as well.
}
\end{center}
\end{figure*}

\begin{figure}
\begin{center}
\psfig{figure=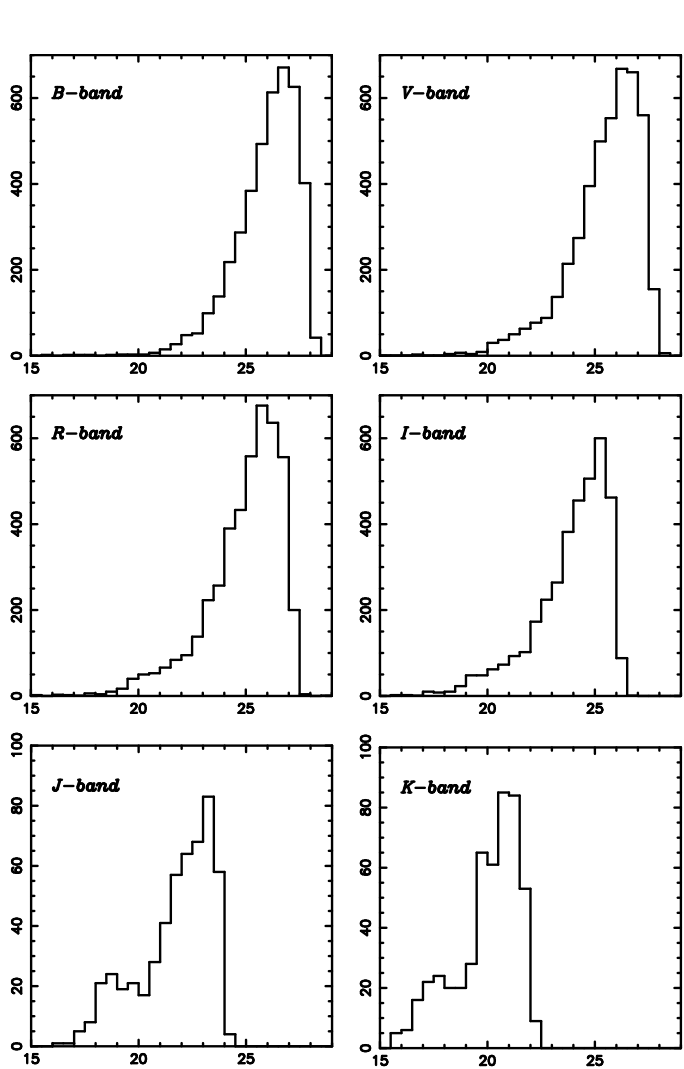,width=9cm}
\caption{\label{magdist} Histograms of the magnitude distribution of the
MS 1008-1224 field : B (top left), V (top right), R (middle left), I (middle 
right), J (bottom left), K (bottom right)}
\end{center}
\end{figure}


\begin{table}
\begin{center}
\begin{tabular}{|c|c|c|c|c|}
\hline
Filter & Exp. Time & Seeing & SB-Lim &  Scale \\
 & (sec.) & (arcsec) &  &  (arcsec/pix) \\
\hline
B & 4950 & 0.72 & 28.25 &  0.2\\
V & 5400 & 0.65 & 27.90 &  0.2\\
R & 5400 & 0.64 & 27.44 &  0.2\\
I & 4050 & 0.55 & 26.37 &  0.2\\
J & 2880 & 0.68 & - &  0.147\\
K & 3600 & 0.45 & - &  0.147\\
\hline
\end{tabular}
\caption{\label{vltsummary} Summary of the characteristics of the VLT 
observations of MS 1008-1224 with FORS and ISAAC. These details are taken 
from the ESO web site. The seeing is the FWHM we measured on the 
co-added image with SExtractor and IMCAT softwares.}
\end{center}
\end{table}


\begin{figure}
\begin{center}
\psfig{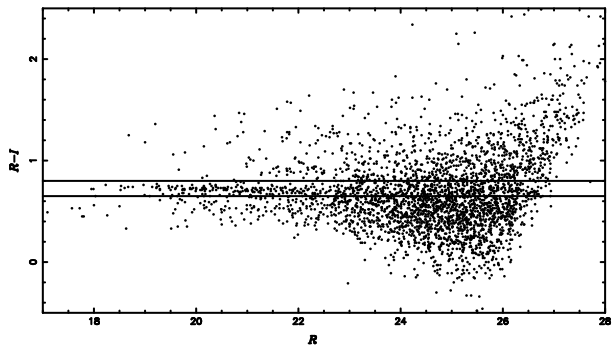}
\caption{\label{colourmag} A colour - magnitude (R-I $v/s$ R) plot for the 
field of MS 1008-1224. The cluster sequence formed by galaxies belonging to
MS 1008-1224 is clearly visible as a horizontal strip 
 at R-I = 0.69 $\pm$0.15.}
\end{center}
\end{figure}


\subsection{Distribution of cluster galaxies}
The cluster 
members were identified from the cluster sequence on the color - magnitude 
plot (R-I $v/s$ R; Fig. \ref{colourmag}). The sequence is almost horizontal 
with these filters. We selected as cluster members all sources having 
$R - I$ = 0.69 $\pm$ 0.15 and R $<$ 24.

The number density and luminosity density distributions are shown in Fig. 
\ref{clusterdensity}. The density at
a point was calculated by computing the area encompassing a fixed number of 
nearest sources as well as by counting the number of sources within a certain
radius around the point. The distributions shown in Fig. \ref{clusterdensity} 
were obtained from counting sources within a fixed radius of 1 arcminute
followed by a smoothing of the resulting density field by a Gaussian of 
0\farcm66 arcminute FWHM. There was no significant change when the cluster 
sequence was extended to a limiting magnitude of R = 27.


\begin{figure}
\begin{center}
\caption{\label{clusterdensity}Number density contours (top) and R-band 
luminosity density contours of cluster members in the fields of  MS 1008-1224.
}
\end{center}
\end{figure}


The galaxy numbers and the light both extend north from the cD galaxy with a 
strong concentration around it. One can discern four prominent peaks in the 
light distribution at (x$_{pixel}$, y$_{pixel}$) $\equiv$ (1100,1100),
corresponding to the location of the cD galaxy, (1040,1400), (1900,1200) and 
(1280,1520). The number distribution also shows all these peaks which suggests 
that the enhancement of light is due to a coherent substructure rather than a 
single bright galaxy. This also indicates that MS 1008-1224 is not yet 
dynamically relaxed.

\subsection{Redshift distribution through the lens}
The deep observations of the cluster field in BVRIJK bands with FORS and ISAAC 
facilitated the determination of the redshift distribution of galaxies with a 
good amount of accuracy using the photometric redshift technique. For our 
purpose, we extrapolated the redshift distribution of background sources within
the ISAAC field to the rest of the FORS field, where J and K band data are 
missing, in order to scale the mass obtained from gravitational lensing
analysis.

\begin{figure}
\begin{center}
\psfig{figure=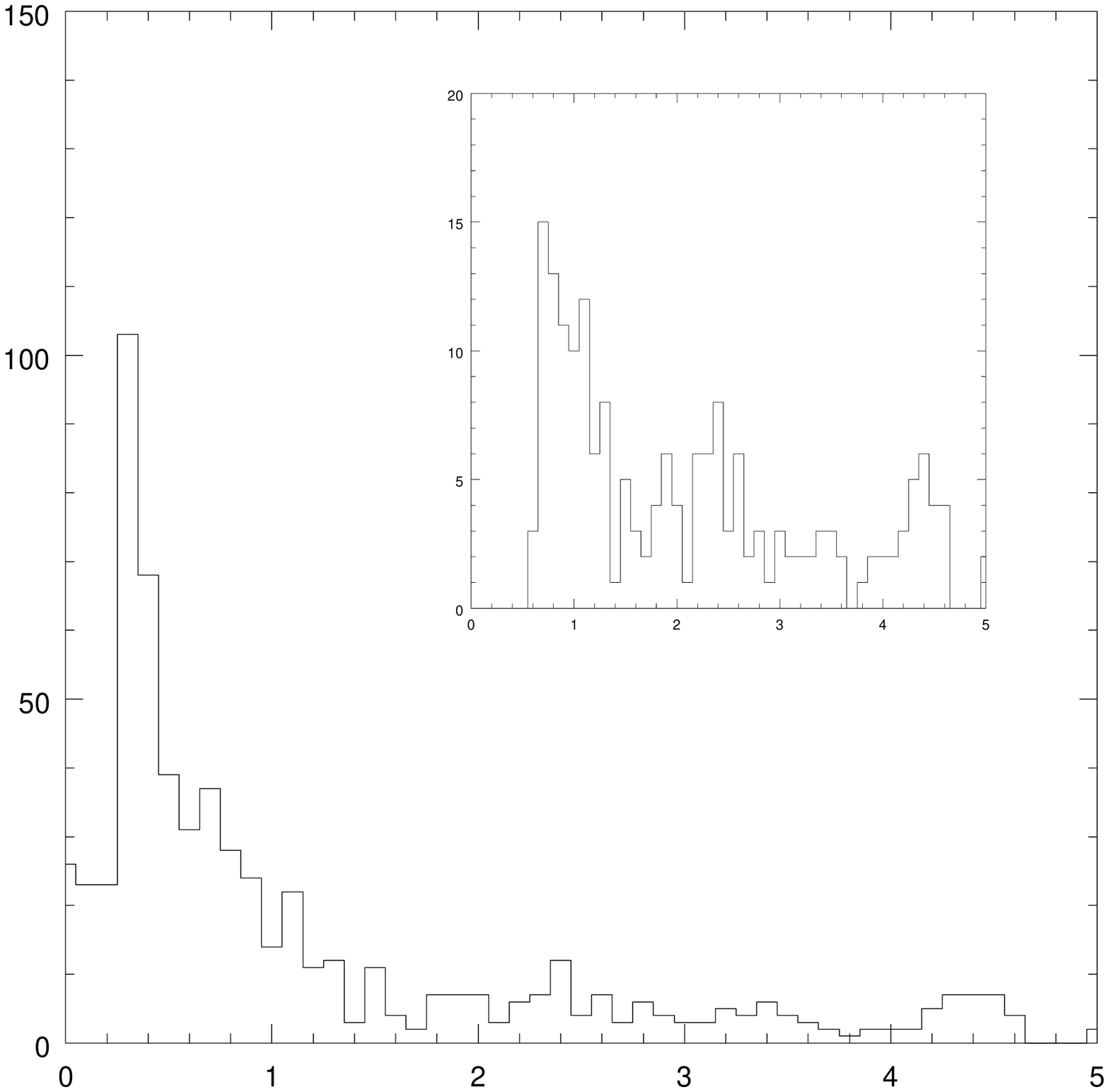,width=9cm}
\caption{\label{zphotdist}Redshift distribution of the galaxies inside the
ISAAC field obtained from the BVRIJ and K data. The peak corresponds to
the cluster MS 1088-1224.  The histogram
inside represents the galaxies having very good measurements
of photometric redshift, in the magnitude range I=22.5-25.5
corresponding to the galaxies selected for the mass reconstruction in
I-band, and with photometric redshift securely larger than the cluster
 (we choose $z \ge 0.4$).
}
\end{center}
\end{figure}

The photometric redshifts were measured using the algorithm described by Pell\'o
et al (1999). It uses photometric data to produce an observed spectral energy
distribution (SED) which is then compared to a set of templates comprising a 
broad range of SEDs of galaxies from the Bruzual \& Charlot evolutionary models
(Bruzual \& Charlot 1993), including a wide range of age and metallicity.
The most probable redshifts were then inferred from a $\chi^2$ minimization.
Since the SED is dependent on many parameters it is important to sample it at 
many points over as large a wavelength range as possible and to get accurate 
photometric measurements. The deep FORS and ISAAC images of MS 1008-1224 are 
therefore ideal for this study. Fig. \ref{zphotdist} shows the redshift 
distributions of galaxies in the ISAAC field; only those galaxies which fit the
templates with $\chi^2 < 1$ have been used. The peak at redshift 0.3 
corresponding to the cluster is obvious.  However, the uncertainties in the 
photometric redshift estimation have smeared this peak between $z$= 0.27 and
$z=$0.40. 

In the depletion analysis (described later) sources at z $>$ 0.4 formed the 
background sample while those at z $<$ 0.27 comprised the foreground sample. In 
the shear analysis, where the sources were selected only by their R-magnitude,
the scaling of the mass, including the effect of contamination due to 
cluster/foreground sources, was done using the redshift distribution of sources
in this magnitude range.

\section{Gravitational lensing analysis in MS 1008-1224}

\subsection{Mass reconstruction from weak shear}

The weak distortion of background sources produced by gravitational lenses can 
be used to construct the projected mass distribution of the lens (Schneider,
Ehlers \& Falco 1992, Kaiser 1992, Fort \& Mellier 1994). A considerable amount
of effort has gone into making the mass reconstruction more accurate and robust
(see Mellier 1999 and references therein).

The remarkable quality of the present data on MS 1008-1224, in terms of seeing, 
depth and stability of the FORS instrument (I. Appenzeller, private 
communication) minimize the pitfalls, the critical PSF correction in particular,
which tend to limit the production of stable and reliable mass maps. The mass 
maps for MS 1008-1224 were constructed by two teams working independently, using
two different algorithms, to assess the robustness of the result. 
Both algorithms use the averaged ellipticities of the galaxies to
compute an unbiased estimator of the gravitational shear. They differ 
in the selection of the background galaxies as well as the reconstruction 
scheme used to obtain the projected mass density.

The mass reconstruction algorithms are rather CPU intensive and we had to run
these algorithms over a wide volume of input parameter space to check the
stability of our results. The considerable facilities (speed, memory and disk 
space) of the TERAPIX Data Centre (http://terapix.iap.fr) played an important
and extensive part in this study.\\

\noindent
{\bf Method 1}\\

\noindent
The IMCAT weak-lensing analysis package has been made publicly available at
the URL\\ http://www.ifa.hawaii.edu/$\sim$kaiser
by Kaiser and his collaborators (Kaiser, Squires \& Broadhurst 1995,
Luppino \& Kaiser 1997). The specific version used was the one modified and 
kindly made available to us by Hoekstra (see Hoekstra et al 1998). 
A description of the analysis  including measurement of the galaxy polarization,
correction for the smearing and anisotropy of the PSF and the sizes of galaxies 
and the expression for the final shear estimate have already been given by 
Hoekstra et al (1998) and references therein and will not be repeated here. 

The maximum probability algorithm of Squires \& Kaiser (1996), with K = 20 
(number of wave modes) and $\alpha$ = 0.05 (the regularisation parameter), was 
used to reconstruct the mass distribution from the shear 
field. Our analysis differs from that of Hoekstra et al only in the 
weighting of the data at the final stage and is described next.

\paragraph{Error weighting \ : \ }
The contribution of gravitational shear to the polarization of a background 
source is only a small fraction of its intrinsic polarization. Consequently one
has to average the shapes of many (10 -- 20) neighboring sources to estimate 
the gravitational shear at that location. Further, since the noise is much 
larger than the signal in individual sources their shapes have to be 
appropriately weighted to obtain meaningful results. Following Hoekstra et al,
we estimated the average value of the shear at a location by 
$\left<\gamma\right>$ = $\Sigma$(W$_i\gamma_i$)/$\Sigma$W$_i$ with 
$\gamma_i$ being the shear (intrinsic + gravitational) of individual galaxies 
and W$_i$ = G$_i$/$(\delta\gamma_i)^2$ the weight comprising the error on the
shear ($\delta\gamma_i$) and G$_i$, a Gaussian factor depending on the distance 
of the $i$-th galaxy from the location for which the average shear was being 
calculated. The two components of the average shear thus calculated, which is a 
function of position, is then to be summed up in the mass reconstruction 
programme along with appropriate co-efficients to generate the mass map.

While this method works quite well it has the disadvantage in that the error
weighting and the Gaussian smoothing are coupled to each other; any reduction
in the Gaussian smoothing scale reduces the effectiveness of the all important
error weighting of the individual shear values and in the limiting case when
the Gaussian smoothing scale includes just one background source there is no
error weighting at all. So, apart from $\left<\gamma\right>$, we also calculated
the error on it, 
$\delta\left<\gamma\right>$ =  $\Sigma$(G$_i/\delta\gamma_i)^2$ / 
[$\Sigma$(G$_i/(\delta\gamma_i)^2)$ ]$^2$. The summations in the mass 
reconstruction programme were modified to take into account this error on 
average shear as well.

\paragraph{Source selection \ : \ } The source detection was done in the R-band
using both IMCAT and SExtractor. Only those background (to the cluster) sources 
which were detected by IMCAT with a significance $>$ 7 and were also detected 
by SExtractor were used for the lensing analysis. Sources with neighbors closer
than 5 pixels (1 arcsecond) were eliminated to reduce the error in the shape
estimation. With a series of subsamples 
spanning a narrow range of $\delta$R = 0.5mag, we detected using Aperture Mass 
densitometry (described later) a lensing signal for sources in the range 22.5 
$<$ R $<$ 26.5. This constituted our master list of 2550 sources for the mass 
reconstruction analysis. The sources from this master list which were detected 
in the other 3 bands (B $\equiv$ 2080, V $\equiv$ 2423 and I $\equiv$ 2417 
sources) constituted the lensing analysis sample for those bands. 

\paragraph {Quality of the mass reconstruction \ : \ }
There are several ways of estimating the quality of the mass reconstruction : \\
(i) A mass reconstruction after the action of a $curl$ operator on the shear 
field (effectively replacing $\gamma_1$ by $-\gamma_2$ and $\gamma_2$ by 
$\gamma_1$) must produce a structure-less noise map (Kaiser 1995). \\
(ii) The boot-strap method\\
(iii) The location and the intensity of the negative values in the mass 
reconstruction.

The lensing data set comprising an almost identical sample of sources in each of
the 4 bands but obtained under different observing conditions and at different 
times should provide a good check on the fidelity of the mass reconstruction 
algorithm. This is particularly true since the gravitational shear signal is 
such a small fraction of the total noise (intrinsic ellipticity plus 
measurement) on individual objects.

We also made mass maps using various Gaussian smoothing scales, from 0 (i.e. no
smoothing) to 30 arcseconds, to confirm the stability of the reconstruction
within an observing band.\\

\noindent
{\bf Method 2}\\

This method uses the raw IMCAT software from Kaiser's home page (see previous
method) corrected for some minor problems (Erben, private communication). 
Objects with a significance $>$ 7 and with a raw (smoothed) ellipticity smaller 
than $0.6$ were used. This resulted in the rejection of all the galaxies with
a corrected ellitpicity larger than 1. Such anomalously high values for the 
corrected ellipticity occurs when the isotropic PSF correction becomes unstable,
as in the case of the faintest and/or smallest galaxies. The mass reconstruction
was done using the maximum likelihood estimator developed by Bartelmann et al. 
(1996) with the finite difference scheme described in Appendix B of Van 
Waerbeke, Bernardeau \& Mellier (1999). The mass reconstruction was not 
regularised and hence the resulting mass maps contain
more noise features than in maps obtained from Method 1. However, as pointed out in
Van Waerbeke et al (1999) and Van Waerbeke (1999) the advantage of method 2 is
that the noise can be described analytically in the weak lensing approximation.
We used this property to derive the significance of the offset between the mass
peak and the cD galaxy.

\paragraph{Noise properties \ : \ }

The galaxy ellipticities were smoothed with a gaussian window

\begin{equation}
W(\thetag)={1\over \pi \theta_c^2} \exp\left(-{|\thetag|^2\over \theta_c^2}\right).
\end{equation}
The noise in the reconstructed mass map is then a 2-D gaussian random field 
fully specified by the noise correlation function (Van Waerbeke 1999):

\begin{equation}
\langle N(\thetag)N(\thetag')\rangle={\sigma_\epsilon^2\over 2} {1\over 2\pi\theta_c^2 n_g}
\exp\left(-{|\thetag-\thetag'|^2\over \theta_c^2}\right).
\end{equation}
%
%

\begin{figure*}
\begin{center}
\caption{\label{rmamass150}Mass reconstruction of MS 1008-1224 using the 
algorithm of Squires \& Kaiser (1996) and a
Gaussian smoothing of 30 arcseconds. The different maps are : B-band (top-left),
V-band (top-right), R-band (bottom-left) and I-band (bottom-right).
The mass reconstructions used sources in the range 22.5 $<$ R $<$ 26.5.
The contours are of equal magnification ($\kappa$) and the levels are 0, 0.02,
0.04, 0.06, 0.08, $\ldots$
The cross-wire in each panel marks the location of the cD
galaxy at pixel (x, y) $\equiv$ (1100, 1100). 
}
\end{center}
\end{figure*}

\subsubsection{Distribution of the dark matter in MS 1008-1224}

The mass reconstructions from B, V, R and I data using the first method and a 
Gaussian smoothing scale of 30 arcseconds are shown in Fig. \ref{rmamass150}. 
Figure \ref{rmamass75}
shows mass reconstructions using the first method with a Gaussian smoothing of 
15 arcseconds (B, V, R and I), and for comparison, Figure \ref{ludomass100}
shows a reconstruction of the I-band image with the second method and
a smoothing of $20$ arcseconds. Figure \ref{highres} is a high resolution, 
though noisy, reconstruction in the four bands using the second method.
The cross-wire on the maps (Figures \ref{rmamass150}, \ref{rmamass75}
and \ref{highres}) indicates the location of the cD galaxy for reference. 
Figure \ref{ludomass100} shows the I band image in the background.

\paragraph{Mass maps with 30 arcseconds smoothing scale}
The reconstructions shown in Fig. \ref{rmamass150} are very similar in shape as 
well as magnitude of the peak ($<$ 5 per cent variation). This certainly speaks 
for the stability of the mass inversion algorithm and the accuracy of the PSF
correction scheme since the PSFs vary from band to band. However, there is 
still the matter of systematic effects in the mass reconstruction. This is 
particularly important in the case of algorithms which use an FFT, like the 
maximum probability algorithm. FFTs often result in correlated noise and 
artifacts which mimic formally significant (i.e. high signal-to-noise) 
structures. For example, it is not clear if the low level spur to the north-east
is real or an artifact. In particular, we can expect the two large masks on the 
MS 1008-1224 CCD images (see Fig. \ref{ludomass100}) to contribute a significant
amount of artifacts to mass 
reconstructions. Eliminating such artifacts will require a deconvolution 
from the mass map of the effect of the (incomplete) sampling of the data plane. 
We checked that this mass reconstruction was reliable by comparing it with mass 
maps obtained using the second method (which does not use an FFT). We found 
that the results of the two methods matched quite well.
It must be noted that the shear inversion algorithms, source 
selection and the redshift range are all different for the two methods and yet 
we obtain similar results. There appears to be an offset between the peak of the
mass map and the cD galaxy which is consistent with the X-ray observations of
Lewis (1999) as well. We have discussed this offset in more detail later.


\begin{figure*}
\begin{center}
\caption{\label{rmamass75}Mass reconstruction of MS 1008-1224 with a smoothing 
scale of 15 arcseconds. The maps are : B-band (top-left), V-band (top-right), 
R-band (mid-left), I-band (mid-right), the intersection of the contours in the 
above four maps (bottom-right) and the noise map (bottom-right). The contours of
convergence ($\kappa$) are 0.0, 0.05, 0.10, 0.15, 0.20, 0.25, 0.30 and 0.35. The
negative (dashed) contours of the noise map have the same spacing. The {\em 
Intersection} image delineates features common to all the bands and are 
therefore most likely to be real. The concentric circles spaced 0\farcm4
apart on the bottom-left panel are the apertures used for the radial profiles of
shear, $zeta$-statistics and mass. The white spot at the centre marks the mass
peak in the low resolution map of Fig. \ref{rmamass150}.
The noise map is mostly between $\pm$0.05 and 
is indicative of the strength of spurious features in the mass maps. The two 
strong features seen around (300, 1700) and (1500, 1800) correspond to the 
location of the CCD masks and are a result of the consequent holes in the data. 
However, the influence of these masks is not significant elsewhere. 
}
\end{center}
\end{figure*}


\begin{figure}
\begin{center}
\caption{\label{ludomass100}Mass distribution of MS 1008-1224 using the 
algorithm of van Waerbeke et al (1999) and a gaussian smoothing of 20
arcseconds  superposed on the R-band image obtained with the FORS on the VLT. 
This map is to be compared with the I-band map of Fig. \ref{rmamass75}, although
it is more noisy here since the reconstruction is not regularised. Despite
the different mass reconstruction algorithm and different source selection
(I-band selected with 22.5 $<$ I $<$ 25.5) the agreement is rather good and
even better when compared to the {\it intersection} image on
Fig. \ref{rmamass75}. The main cluster mass
is oriented north-south and offset to the north of the CD. The extensions
around the main clump in the {\it intersection} image in Fig. \ref{rmamass75}
seem to coincide with the clumps around the cluster here.
The blank regions in the top corners are the CCD masks used
to blank out bright stars.
}
\end{center}
\end{figure}

\begin{figure}
\begin{center}
\psfig{figure=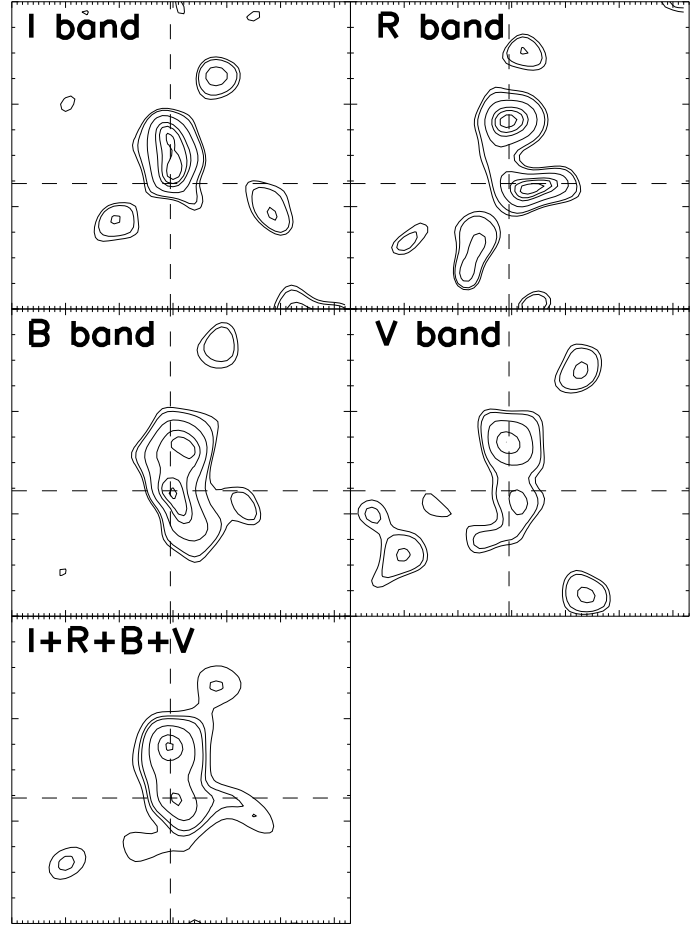,width=9.cm}
\caption{\label{highres}High resolution mass reconstruction of MS 1008-1224 
with a smoothing scale of 15 arcseconds using Method 2. The plot at the 
bottom-left is the average of the reconstructions in the four bands. The noise 
is reduced by a factor of two in this average map, and it shows the significant 
elongation of the mass distribution toward the north, as well as a double peak 
internal structure. The dashed cross shows the position of the cD galaxy which
coincides with the south extension of the mass profile.
}
\end{center}
\end{figure}

\paragraph{Noise map}
The bottom-right panel of Fig. \ref{rmamass75} shows a mass map obtained from
the $curl$ of the shear field. In the absence of systematics this is expected to
be a featureless image. This is clearly the case over much of the field with the
intensity being in the range $\pm$0.05. However, we also see strong positive and
negative features at around (300, 1700) and (1500, 1800) which coincides with
the location of the CCD masks reported earlier. It is heartening that the
influence of the large holes in the data, as a consequence of the masks, does
not extend over the rest of the field.

\paragraph{Mass maps with higher resolution }
We also constructed mass maps using Gaussian smoothing scales of 0 to 25 
arcseconds to (i) explore the mass distribution in more detail, (ii) to
confirm that the maps from different scales are consistent with each other
and (iii) eventually to probe the inner structure of the cluster.
Naturally, the images produced with smaller smoothing scales are noisier. We
found that the 15 arcseconds smoothing produced all the details that we could
reliably extract and so we have shown these (from B, V, R and I data) in
Fig. \ref{rmamass75}. Considering its noisy nature, we have also reproduced
in the same figure an {\em intersection} map, comprising the intersection of
the contours from all the four bands to delineate the most stable and hence
presumably the more genuine structures (as against artifacts). The general 
features on Fig. \ref{rmamass75} are remarkably stable and consistent with the 
low resolution map on Fig. \ref{rmamass150}. 
The inner mass concentration is elongated to the north and consists of 2 peaks. 
Figure \ref{highres} shows maps produced using the second method (which is not 
regularised) with a smoothing of 15 arcseconds in each of the 4 bands. For 
convenience the displayed field is restricted to the inner 
3\farcm3$\times$3\farcm field. Though noisy, all of them do show the double
structure delineated by the reconstruction using the first method. In fact,
the co-added map included in the same figure has considerably less noise
and clearly shows the double structure. Thus, all the high resolution images 
show that the central mass distribution of MS 1008-1224 is consists of a 
double structure. The northern, and the more dominant peak, is coincident with
the excess of galaxies seen in Figure \ref{clusterdensity}. The southern peak
is well localised on the cD galaxy.

\paragraph{Offset significance}

From the noise properties described in Section 3.1 it is possible to obtain the 
dispersion of the peak mass position due to the random intrinsic ellipticities 
and the random position of the background galaxies. The best way of measuring 
the significance of the offset would have been a parametric model for the mass 
distribution using a parametric bootstrap method to generate a large number of
mass reconstructions with different noise realisations and measuring the
dispersion of the centroid in these various realisations. Unfortunately such a
parametric model is not available and the best that we can do is to consider the
mass map reconstructed from the data itself as the mass model. Using the noise 
model in Section 3.1 we can then generate different noise realisations (since 
each realisation consists of adding a gaussian correlated random noise to the 
reconstructed mass map) which {\it simulates} mass reconstructions with 
different realisations of the intrinsic ellipticities and different galaxy 
positions.

\begin{figure}
\begin{center}
\psfig{figure=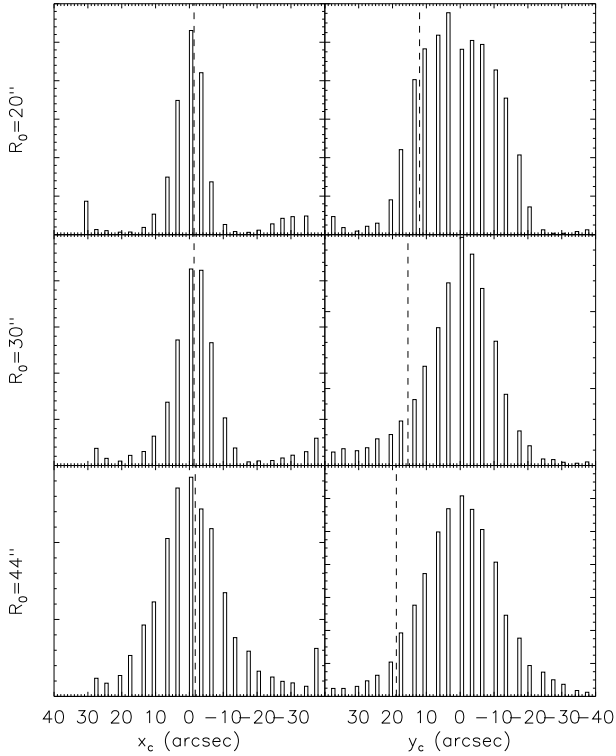,width=8.cm}
\caption{\label{centroide}Parametric bootstrap resampling of the centroid of 
the mass clump on the I-band image. The left column corresponds to the X-axis 
offset, and the right column to the Y-axis offset. The position of the cD galaxy
is indicated by the vertical dashed lines. Three different scales are 
represented. The bootstrap were done over 5000 realisations of the noise using 
the noise model of Van Waerbeke (1999) (see Section 3.1). It shows that the 
offset is not statistically significant.
}
\end{center}
\end{figure}

We generated 5000 noise realisations for three different smoothing scales 
$R_0=20", 30", 44"$ with the I-band image. For each realisation we identified 
the peak location, and measured the dispersion among the 5000 realisations for 
the X and Y directions. The results are shown in Figure \ref{centroide}. They
show that the offset is most (but not very) significant along the y-axis and in
the lowest resolution image which strengthenes the idea that the offset seen 
at low resolution is not real and is a consequence of 2 mass components in the
central regions of MS 1008-1224.

\paragraph{Noise map}
The bottom-right panel of Fig. \ref{rmamass75} shows a mass map obtained from 
the $curl$ of the shear field. In the absence of systematics this is expected to
be a featureless image. This is clearly the case over much of the field with
the intensity being in the range $\pm$0.05. However, we also see strong positive and 
negative features at around (300, 1700) and (1500, 1800) which coincides with
the location of the CCD masks reported earlier (see also the Figure
\ref{ludomass100} which shows the masks and the induced artefacts
in the mass map). It is heartening that the
influence of the large holes in the data, as a consequence of the masks, does
not extend over the rest of the field.

\paragraph{A summary of the Features in the mass reconstruction \ : \ }
Having dealt with the principal pitfalls of the mass reconstruction algorithms
and the measure procedures, we now proceed to 
point out the more reliable features in the mass distribution : 

\begin{itemize}

\item The mass is concentrated in the vicinity of the cD and extends towards the
north. This is clearly seen in the high resolution images, particularly in the
image formed by the intersection of the mass maps in the 4 bands and in the
coadded high resolution image.

\item The central concentration when viewed at higher resolution consists of
two components. The north-east component appears to be stronger than the 
south-west peak with which the cD galaxy coincides. However, the significance of
the difference in relative strengths is marginal.

\item An analysis of the stability of the peaks in the mass reconstruction
indicates that while the cD appears to be associated with the southern mass
component, its offset from the centre of mass is not highly significant.

\item There are other weaker features seen in the high resolution image. 
For most of them it is not possible to ensure that they are real
mass overdensities because they are dangerously close to the edge or the
masks of the CCD. One of these, the south-west extension
is seen in all the maps, both high and low resolution. The spur extending to the
north-east (close to the mask) is also seen in the noise map
(Fig. \ref{rmamass75}) and is therefore likely
to be an artifact. We will come back to these later when we compare the mass 
structures with optical and x-ray data.

\end{itemize}

\subsubsection {Mass Profile from tangential shear}

The mass from weak shear, $M_{shear}$, may be obtained from Aperture Mass
Densitometry or the $\zeta$-statistics described by Fahlman et al (1994) and
Squires \& Kaiser (1996). In brief, the average tangential shear in an annulus
is a measure of the average density contrast between the annulus and the region
interior to it. i.e. The average magnification $\kappa$ ($\equiv \Sigma /
\Sigma_{cr}$), the ratio of the surface mass density to the critical surface
mass density for lensing, as a function of the radial distance $\theta$ is given
by

\begin{eqnarray}
\kappa\left(<\theta_1\right)& =& 
\kappa\left(\theta_1<\theta<\theta_2\right) \nonumber \\
 & &\displaystyle{  +
\frac{2}{1 - \left(\frac{\theta_1}{\theta_2}\right)^2}
\int_{\theta_1}^{\theta_2} \left<\gamma_t(\theta)\right> {\rm d} \left(\ln\theta\right)} \ ,
\end{eqnarray}

\noindent
where, $ \gamma_t = - (\gamma_1\cos{2\psi} + \gamma_2\sin{2\psi}) $
is the tangential component of the shear with $\psi$ being the angle between the
position vector of the object and the x-axis. One can get an estimate of the
error on $\gamma_t$ by calculating
$\left<\gamma_r\right> =
-\left<-\gamma_2\cos{2\psi} + \gamma_1\sin{2\psi}\right> $
which is a measure of the random component of the galaxy shear due to intrinsic
ellipticity and observation noise and is expected to be zero around any closed
strip. From this we can
obtain the average magnification within a series of apertures of radii
$\theta_i$, $i = 1\ldots n$, to derive the radial profile

\begin{eqnarray}
\kappa\left(<\theta_i\right)& = &
\kappa \left(\theta_n<\theta<\theta_b\right) 
\nonumber \\
 & & +
\frac{2}{1 - \left(\frac{\theta_n}{\theta_b}\right)^2}
\int_{\theta_n}^{\theta_b} \left<\gamma_t\left(\theta\right)\right> {\rm d
}\left(\ln\theta\right) \nonumber \\
 & & + 2 \int_{\theta_i}^{\theta_n} \left<\gamma_t\left(\theta\right)\right> 
 {\rm d} \left(\ln\theta\right),
\end{eqnarray}

\noindent
where, the region $[\theta_n, \theta_b]$ is the boundary annulus which provides
the reference density (the first term) in excess of which the interior density
values are obtained. Thus this method provides only a lower limit to the
lensing mass estimate. It is to be noted that this expression has been cast such
that the first term, which cannot be calculated and hence is to be neglected, is
the average within the annulus [$\theta_n, \theta_b$] and not the average
density interior to $\theta_b$. Therefore, the effect of neglecting this term
will be quite small if the data extend sufficiently far from the cluster
centre. \\
The mass within an aperture is given by

\begin{eqnarray}
M (<\theta_i) & =& \kappa(<\theta_i) \ \Sigma_{cr} \cdot \pi \left(\theta_i D_{ol}\right)^2 \nonumber \\
 & = & \kappa\left(<\theta_i\right) \ \theta_i^2 \,\frac{c^2}{4G} 
\left<\frac{D_{ls}}{D_{os}D_{ol}}\right>^{-1}
\end{eqnarray}

\noindent
where $D$ is the angular size distance and its subscripts, $l$, $o$, and $s$,
refer respectively to the lensing cluster ($z = 0.3062$), observer ($z = 0$) and
the background lensed sources ($z = z_s$). As has been explained earlier this 
high quality dataset allows us to estimate photometric redshifts of sources in 
the ISAAC field. Assuming that the ISAAC field is
representative of the FORS field as a whole we have good photometric redshifts
($\chi^2 < 1$) for 356 objects with 22.5 $<$ R $<$ 26.5 (the lensing analysis
range). Of these 302 lie behind the cluster at z $>$ 0.31. We find
$\displaystyle{ \left<\frac{D_{ls}}{D_{os}D_{ol}}\right>^{-1}}$ = 1.383$\pm$0.031 Gpc 
which also includes a correction for the dilution of the lensing signal due to
foreground objects being in the selected magnitude range. The mass is therefore
given by

\begin{equation}
M(<\theta_i) = 6.14 \times 10^{14} h^{-1}{\rm M_{\odot}}\ 
 \kappa(<\theta_i) \ {\theta_i^{'}}^2 ,
\end{equation}
where $\theta_i^{'}$ is in arc-minute.  

\begin{figure}
\begin{center}
\psfig{figure=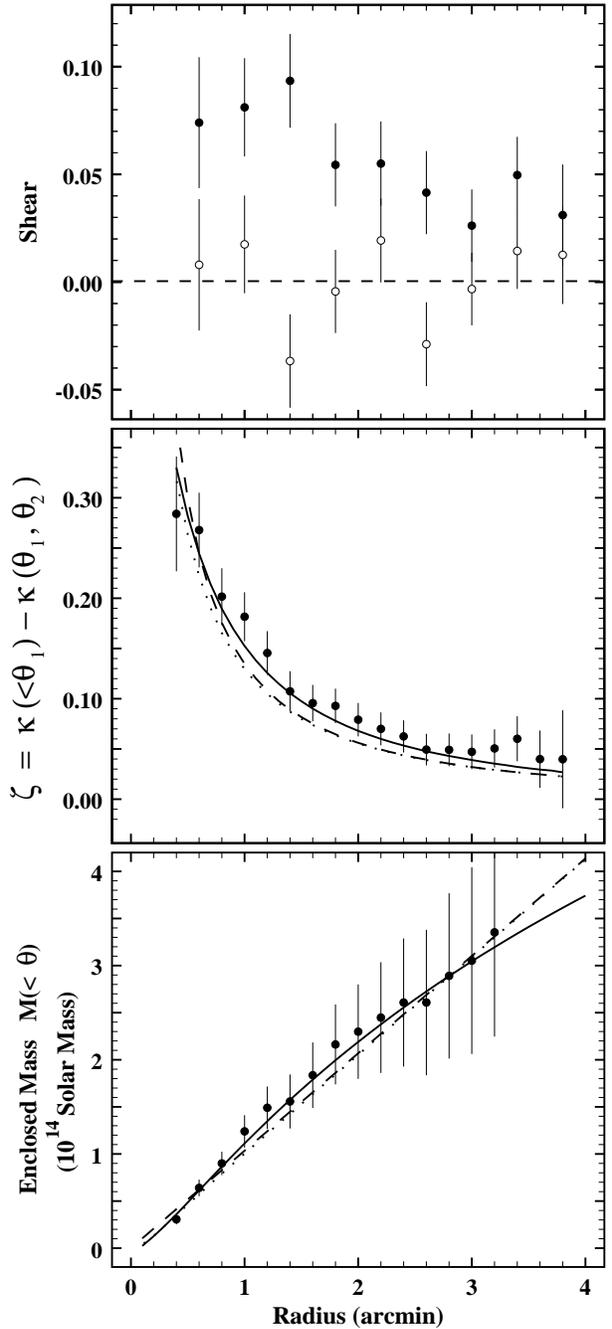,width=8.cm}
\caption{\label{radialprofile} Radial profiles in MS 1008-1224 with the
weak lensing analysis of the  $R$-selected galaxies.
The top panel is the radial tangential shear, centered around the
central position given by the joined analysis from weak lensing and
depletion (1150,1150).
The filled circles are the measured tangential shear. The open circle are
the data dat where $\gamma_1 \rightarrow \gamma_2$,
and $\gamma_2 \rightarrow -\gamma_1$. The transformated data do have zero
amplitude, expected if the tangential shear were produced by a
gravitational lensing effect.  The bottom panel is the mass enclosed
within circular radii. We mass profiles were fitted by an isothermal
sphere with a core radius (dashed lines) and an NFW profile (lull
lines). The curves show that the NFW profile is slightly better than 
the isothermal sphere, but the difference is small.
}
\end{center}
\end{figure}

The radial profile of the shear, the $\zeta$-statistic, 
$\zeta = \kappa (<\theta_1) - \kappa(\theta_1, \theta_2)$ , 
and the mass are shown in Fig. \ref{radialprofile}. The apertures
were centered on the centroid of the mass distribution (x$_{pixel}$, y$_{pixel}$)
$\equiv$ (1175, 1175) in the low resolution mass map. The centre and the 
apertures, spaced 120 pixel = 0\farcm4 apart, are marked for reference
on the mass map in the bottom-left panel of Fig. \ref{rmamass75}. The annulus 
between 3\farcm2 and 3\farcm6 radii was used as the boundary annulus to set 
the zero of the density scale.

The upper panel of Fig. \ref{radialprofile} shows tangential shear (filled 
circles) within an annulus centered on the radius. Also plotted in the same 
panel are the values of $\gamma_r$ (open circles) which are distributed around
zero as expected for a signal due to gravitational lensing. The middle panel 
shows the $\zeta$-statistics which is the best expression for comparing the 
data with model fits. The bottom panel shows the mass derived from the shear 
data. It must be noted that the shear points (top panel) are independent of 
each other while the points in the lower two panels are not independent; 
every point makes use of all the shear data exterior to it.

The bars denote $\pm1\sigma$ error in all the panels. We expect that the 
quoted errors are actually overestimates because of the linear morphology of 
the mass and the presence of the two large masks. Since the mass extends to the 
northern boundary of the field of view, it is likely that we are underestimating
the mass of this cluster.  We have confirmed that the error estimated
for the individual shear values and used in the weighting is appropriate by
checking that $\sqrt{\left<\gamma^2_r\right>/N}  \ 
\simeq (\Sigma^{^N}_{_1} 1/(\delta\gamma_r)^2)^{-1} \
\simeq (\Sigma^{^N}_{_1} 1/(\delta\gamma_t)^2)^{-1} $, where $N$ is the total 
number of lensed galaxies in the shell.

\subsection{Magnification bias and Radial Depletion}

As a result of the combined effect of deflection and magnification of light, 
the number density of galaxies seen through the lensing cluster is modified. In 
the case of a circular lens, the galaxy count at a radius $\theta$ is
\begin{equation}
N(<m,\theta)=N_0(<m) \mu(\theta)^{2.5\alpha-1} \ ,
\end{equation}
where $\alpha$ is the intrinsic (without lensing)
slope of the galaxy count, $\mu$ the gravitational magnification, and
$N_0$ the intrinsic galaxy number density (hereafter, the zero-point).  
Depending on the value of 
$\alpha$, we may observe an increase or a decrease in the number of galaxies
in the central region out to a limiting radius which depends on the shape of
the gravitational potential and the redshift distribution of the background 
sources.  This effect produced by the  magnification bias has already been 
observed in some lensing clusters (Broadhurst et al 1995, Fort, Mellier \&
Dantel-Fort 1996, Taylor et al 1998, Broadhurst 1998). It is particularly 
obvious in very deep observations as the slope of galaxy counts decreases 
to values as low as $\alpha \approx 0.2$ at faint levels. The very deep FORS and
ISAAC observations of MS 1008-1224 are therefore well suited for studying this 
effect. In particular, the location of the depleted area provides an independent
measurement of the position of the centre of the cluster mass distribution.

\subsubsection{Evidence of depletion of lensed sources}

As described earlier, we considered as foreground those galaxies which were at
$z_{phot}<0.27$ and as background (lensed) those at $z_{phot}>0.4$. We minimized
misclassification by considering only those which had a photometric redshift
solution with $\chi^2 < 1$. In order to be consistent with the shear analysis
samples, we focussed only on the $R[22.5-26.5]$ (method 1) and the 
$I[22.5-25.5]$ (method 2) samples.
We inferred from the FORS data that the slope of the galaxy counts of these two 
samples were $0.192$ and $0.233$, respectively, which are significantly below 
0.4. Therefore, we expected significant depletion with a clear indication of the
location of the centre of the mass distribution.

\begin{figure}
\begin{center}
\psfig{figure=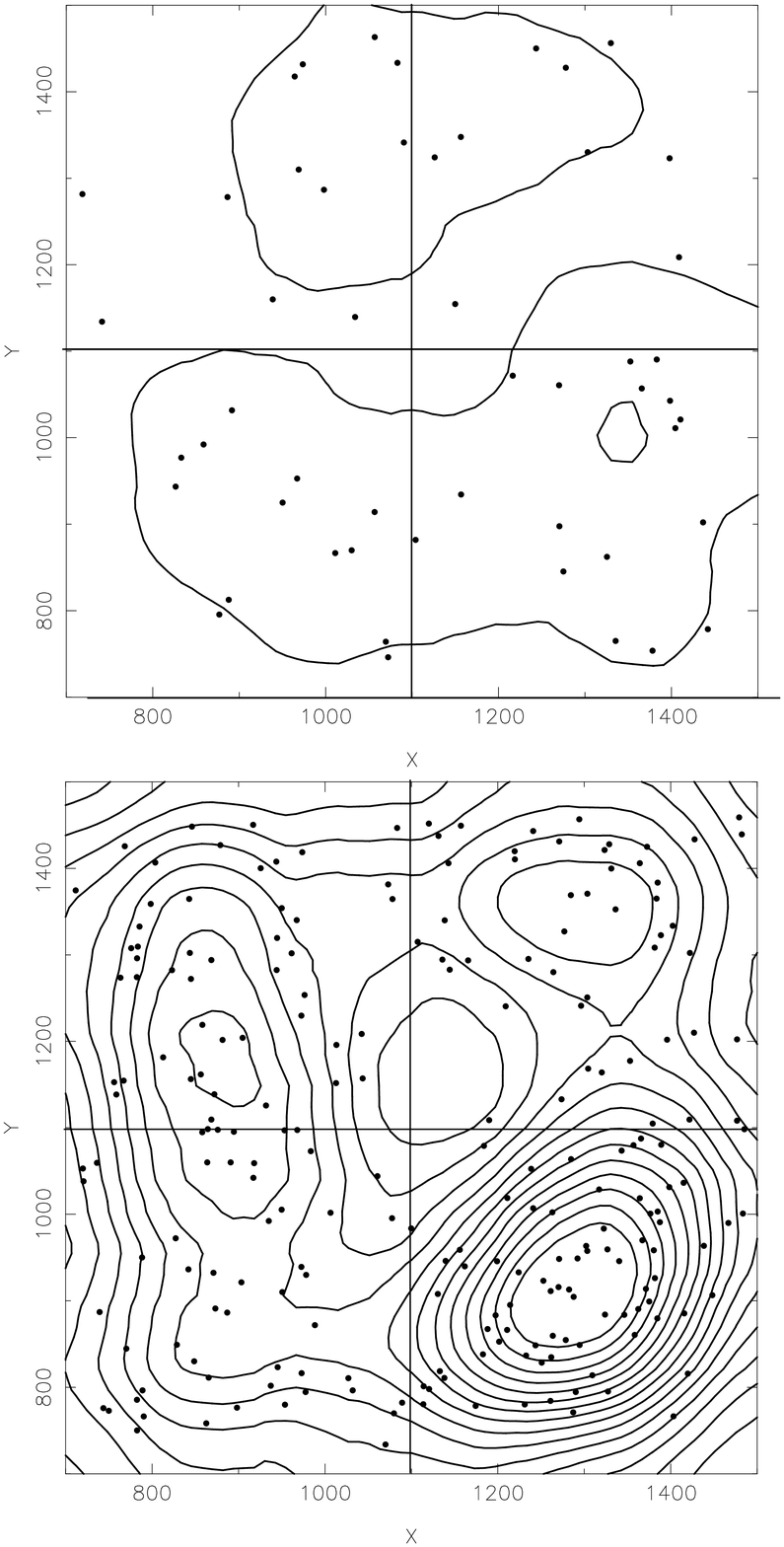,width=8.cm}
\caption{\label{depletion} Number density contours of the ISAAC field.
The contours show the galaxy number density of the sample 
 $I[22.5-25.5]$ having good 
photometric redshift.   The points are the positions of each 
 galaxy of the samples. The top panel is the foreground sample
($z<0.27$), and the bottom panel the background (lensed) galaxies ($z>0.4$).
  The distribution of the foreground is, as expected, almost uniform. 
 In contrast, the lensed galaxies are not uniformly distributed and the
depletion area is clearly visible. The central contour is 
 centered around the position 
 of the dip.  The cross-wire marks the position of the cD. An offset
(15 arcseconds west and 20 arcseconds north) is
clearly detectable between the maximum depletion location and the cD galaxy,
confirming the results obtained previously with the weak shear 
analysis. This is the first evidence for 
magnification bias based on sources with known redshift.
Note the strong excess of
galaxies in the South-West quadrant (quadrant $Q4$), which seems to
result from a cluster of galaxies at very large redshift which is lensed
 by MS 1008-1224.
}
\end{center}
\end{figure}


Figure \ref{depletion} shows the projected number density of galaxies having 
good photometric redshifts from BVRIJK data and in the magnitude range 
$I[22.5-25.5]$.  Galaxies at $z >$ 0.4 (background, lensed) are on the right. 
The control sample of galaxies at $z <$ 0.27 (foreground) are on the left.
The difference between the two panels is clear : the foreground distribution is 
essentially random; in contrast, a strongly depleted area is visible near 
the centre of the field in the background distribution. This is the first 
evidence of the magnification bias effect based on a large sample of background 
galaxies with a known redshift distribution. The cross-wire marks
the location of the cD galaxy. The offset of about 20 arcseconds north and 15 arcsecond west between the cD and the depletion is consistent with the
previous shear analysis. The depletion and its offset from the
cD are also seen in the R-selected sample.

\subsubsection{Mass profile from magnification bias} 

The modification of the radial distribution of galaxy counts due to the 
magnification bias probes the dark matter distribution. In the weak lensing
regime the relation simplifies to :
\begin{eqnarray}
N(<m,\theta)&=&N_0(<m) \mu(\theta)^{2.5\alpha-1}\\
 &\approx& N_0(<m)\left[1+2\kappa\left(\theta\right)\right]^{2.5\alpha-1}  . 
\end{eqnarray}
Hence the convergence $\kappa(\theta)$  is
\begin{equation}
\kappa(\theta)={1 \over 2} \left[\left(N(\theta) \over N_0 \right)^{1 \over 2.5
\alpha -1} -1 \right] \ ,
\end{equation}
and the total mass inside the radius $\theta$ (in arcminute) is :
\begin{equation}
M(\theta)= 4.4 \times 10^{14} h^{-1} {\rm M_{\odot} }  \left<{D_{ls}
\over D_{os}}\right>^{-1}  \left({D_{ol} 
\over {\rm 1Gpc}}\right) \int_0^{\theta}
\theta \kappa(\theta) {\rm d}\theta \ .
\end{equation}
The slope of the galaxy counts, $\alpha$, 
 is calculated from the deep FORS data. The depletion curves are also
 directly provided by the data, either from FORS or from ISAAC, 
depending on the sample we work on.  The quantity 
$\left<{D_{ls}/ D_{os}}\right>^{-1}$ is computed from the redshift distribution
shown in Figure \ref{zphotdist}. So, in principle, we can get the radial mass 
distribution by merely counting galaxies on the FORS/ISAAC data. However, 
measurement of the radial depletion is tricky. Since the field of view of ISAAC 
is smaller than the radial extent of the depletion area, the number density even
at the edge of the ISAAC field was not the appropriate value for the zero-point,
$N_0$.

Even more critical (to the mass estimate) was the significant enhancement of 
galaxy number density to the south-west direction which enhanced the depth of 
the radial depletion. This enhancement is clearly visible on the bottom-right 
quadrant (hereafter the fourth quadrant or Q$_4$) of the right panel of Figure 
\ref{depletion}. A visual inspection of the FORS images shows that many 
faint and distorted galaxies are present between a radius of 250 and 400 
pixels from the centre of depletion. This excess is probably produced by galaxy 
clustering beyond the MS 1008-1224. They are all elongated tangentially with
respect to the centre of depletion, which is expected if the distortion is due
to gravitational shear. Remarkably, the shape of the number density contours of 
the smoothed distribution shows similar ellipticity and orientation as 
individual galaxies, as if they were magnified accordingly. 

\begin{figure}
\begin{center}
\psfig{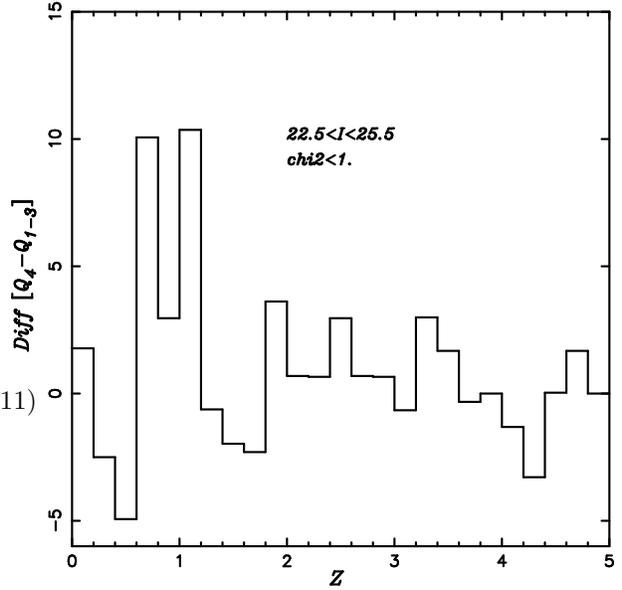}
\caption{\label{diffhistoz} Comparison of the redhsift distribution  of the 
galaxies inside the South-West quadrant ($Q_4$) with respect to the others 
($Q_1+Q_2+Q_3$). 
A strong excess of galaxies shows up, at redshift 0.9.  This excess
 is highly significant, more than $6\sigma$ above the average computed
from the quadrants $Q_{1,3}$. We conclude that a distant cluster of
galaxies lies in this redshift range.
}
\end{center}
\end{figure}

We compared the photometric redshift distribution of galaxies in Q$_4$ to those
in the other three quadrants (hereafter Q$_{1-3}$). The distributions, plotted
in Fig. \ref{diffhistoz}, show a significant excess of galaxies at redshift 0.9.
Quantitatively, in the magnitude range $I[22.5-25.5]$, the number of galaxies 
with photometric redshifts in the range $z = 0.8 - 1.1$ is 61 in $Q_4$, whereas 
the three other quadrants together totalize 38 galaxies. So the significance 
 level of the
excess is $9.5\sigma$.  We therefore conclude that there is  a distant cluster 
of galaxies behind MS 1008 and at $z \approx$ 0.9, which is globally magnified and
sheared according to the magnification factor of each individual galaxy.
This case of cluster-cluster lensing is the first ever observed so far.

\begin{figure}
\begin{center}
\psfig{figure=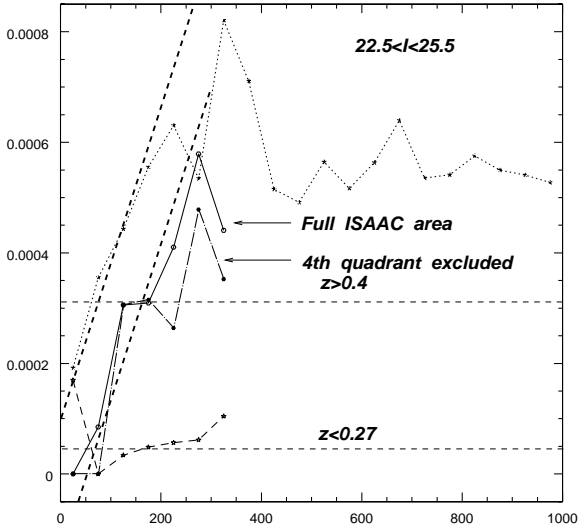,width=8.cm}
\caption{\label{radialdepletion} Galaxy number density as function of the radial
distance with respect to the centre found from the depletion
 (1150,1150). All the plots are for galaxies selected in the 
magnitude range $22.5<I<25.5$. The scale is in pixel units 
 (hence 0.0001 corresponds to 9 gal.arcmin$^{-2}$. 
 The dotted curve (top) is the number 
 density  found in the FORS field. The plateau at large radius provides 
the zero-point for the full sample.  The other curves are 
 computed from the galaxies inside the ISAAC field, having 
 a good photometric redshift.  The full line curve shows
 the galaxy count form galaxies inside the ISAAC field, including 
the bottom-right quadrant where a background  cluster seems present.
 The dot-dashed curve show the same plot, where the bottom-right
quadrant is now excluded.  Finally, he dashed curve (bottom)  
 is the galaxy number density of galaxies with $z_{phot}<0.27$. It
is almost flat since the foreground cannot be depleted.  
The dashed straight lines show the mean number density 
}
\end{center}
\end{figure}

This remarkable lensing event has an unfortunate impact, however, on the 
quantitative analysis of the depletion. This background cluster results in a
"spurious" deepening of the MS 1008-1224 radial depletion profile. 
Furthermore, since this excess is located at the boundary of the ISAAC field, it
is impossible to find out the value of $N_0$ (the background density in the
absence of a lens, usually obtained from near the edge of the field) from the 
ISAAC data alone. We need additional information extrapolated from the FORS 
field as a whole.

We selected galaxies of the FORS field fainter than the brightest cluster 
members and which are outside the 
 $(I-R)-I$ Color-Magnitude strip defined by cluster members. The 
  contamination by background galaxies and residual cluster galaxies 
 should not be a critical issue and more likely only changes the value 
of the minimum, not the shape of the depletion curve. Then, we 
compute the radial galaxy number density of the FORS data within the
ISAAC area which excludes the bottom-right quadrant of Figure 8. 
 Fig. \ref{radialdepletion} shows the growth curves which exclude the forth 
quadrants of the FORS field as well as the ISAAC subsamples
 for which we got photometric redshifts. A  depletion is visible 
as well as a flat distribution at large distance, but beyond the 
 ISAAC field and the position of the background cluster at $z\approx
0.9$. The plateau provides the 
zero points, $N_0^{FORS}$, of all $22.5<I<25.5$ galaxies regardless their 
redshifts. Since the FORS field is composed of foreground galaxies, possibly
 few cluster members and background galaxies, the depletion of the 
 FORS data can be separated in to components, the foreground (unlensed)
galaxies, $Frg$, which include also cluster members, and the background
(lensed) galaxies, $N_0^{FORS(z>0.4)}$, which is composed of all galaxies
with redshift larger than 0.4.  :
\begin{equation}
N = N_0^{FORS(z>0.4)}\mu^{2.5\alpha-1} +Frg \,
\end{equation}
where $\mu$ is the magnification, $\alpha$ the slope of the galaxy
counts in this magnitude range. $N_0^{FORS(z>0.4)}$ is the zero point 
 of the background galaxies; that is the galaxies of the FORS field
having a redshift higher than $z=0.4$.  $Frg$ is the density of foreground 
 (unlensed) galaxies. At large distance, the magnification is negligible
and  we have 
\begin{equation}
N_0^{FORS}=N_0^{FORS(z>0.4)}+Frg
\end{equation}

The analysis of the ISAAC field is more complex. 
 Because the success rate of the photometric redshift 
technique is not 100 per cent of the galaxies having good 
photometric redshifts $z_{phot}>0.4$, the distribution $N(z>0.4)$ 
observed from the photometric redshift of the ISAAC field
  is only  an unknown  fraction $k$ of the 
 galaxies having a redshift larger than 0.4: 
\begin{equation}
N(z>0.4) = k \ N_0^{z>0.4}\mu^{2.5\alpha-1} \ ,
\end{equation}
where $N_0^{z>0.4}$ is the asymptotic value of the galaxy number
density of galaxies with redshift larger than 0.4 \ . At large distance,
$N_0^{z>0.4 }$  is independent of the position, so it must correspond 
 to the background galaxies of the FORS field. Therefore we always have:
\begin{equation}
N_0^{z>0.4}=N_0^{FORS(z>0.4)}
\end{equation}
So the key point is to estimate which fraction $k$ of the background
galaxies have a good photometric redshift and which fraction of the 
 FORS sample is at redshift larger than 0.4. This will provide the
zero-point of our sample.\\
The estimation of the foreground (namely, those having $z<0.4$) 
 turns out to be difficult. We cannot use only the 
sample with photometric redshift $z<0.27$ because we do not know which fraction
of the foreground have a good photometric redshift. From the depletion 
curve of the galaxies with photometric redshift smaller than 0.27, we
 have a lower limit ($Frg>4.5 $ galaxies arcmin${-2}$). However, 
this does not include galaxies with redshift $0.27<z<0.4$. A crude estimate 
can be provided by the depletion curve itself. As we can see on Fig 
\ref{radialdepletion}, the depletion curves of the FORS data is linear   
below $r=200$ pixels. The intersection with the vertical axis provides 
a good idea of the background estimate, 
$Frg \approx 9.0 $ galaxies arcmin$^{-2}$.  
Therefore, we have 
\begin{equation}
N_0^{FORS(z>0.4)}=N_0^{z>0.4}=39.6 \ {\rm gal.arcmin}^{-2} \ .
\end{equation}
We can now estimate the fraction $k$.  From Equations (10), (11) and 
(12) we have
\begin{equation}
k={N(z>0.4) \over  N -Frg} \ .
\end{equation}
The ratios of the galaxy counts can be explored by 
comparing  the depletion
curves of the FORS sample to the depletion curve of the 
 photometric redshift sample inside the 
 ISAAC area, excluding the forth quadrant (see Fig.
\ref{radialdepletion}).  
 Since the depletions probe the same 
parent lensed galaxies, we do expect that the shapes of the 
curves of the FORS sample and the photometric redshift sample 
 should be similar. In principle, it is possible to infer $k$ from 
 the measurement of the ratio given in Eq. (15) at various
points. However, because $k$ depends on the magnification it varies along
 the radial distance. Furthermore, the depletion curve shows strong
fluctuations due to Poisson noise and clustering. In order to minimize
these effects, we locally fitted the curve by a straight line and 
compute and average $k$ inside some radii. 
 Excluding the area $200<r<400$ which shows some residuals of the
distant cluster, we find $<k> = 0.65_{-0.10}^{+0.10}$.  
So, the zero-point of the 
$N(z>0.4)$ sample is $26.3_{-4.5}^{+3.60}$ gal.arcmin$^{-2}$. \\
Using this value we can therefore compute directly $\kappa(r)$.  The 
results are shown on Table 2.
\begin{table*}
\begin{center}
\begin{tabular}{|c|c|c|c|c|}
\hline
$\theta$  & M($\theta,k$=2.42)&M($\theta,k$=2.92) & M($\theta,k$=3.32)&
M($\theta,k$=2.00) Simple offset\\
(arc-min) & $10^{14} h^{-1} {\rm M}_{\odot}$ & $10^{14} h^{-1} {\rm M}_{\odot} $
& $10^{14} h^{-1} {\rm M}_{\odot}$&  $10^{14} h^{-1} {\rm M}_{\odot}$ \\
\hline
0.21 & 1.26 & 2.70 &  4.19 & 0.31\\
0.50& 1.56 & 3.94 &  6.03 & 0.31\\
0.67 &1.56 & 4.03 &  7.04 & 0.31\\
0.83 &1.56 & 4.03 &  7.20 & 0.31\\
1.00 &1.56 & 4.03 &  7.20 & 0.31\\
\hline
\end{tabular}
\caption{Mass from the growth curve of galaxy counts of the 
sample having photometric redshift larger than 0.4. The magnitude range
of the galaxies is $22.5<I<25.5$, and the galaxies of the forth quadrant
which contains a background cluster have been excluded. The mass is
computed for 4 values of $k$. The last one is obtained by simply assuming
the depletion curves are similar and are just offset by an amount
which gives directly the amount of galaxies at redshift larger than 0.4.
 For a given value of $k$, the mass increase until we reach the 
asymptotic value of the zero-point. Due to the poor statistics we cannot
provide an accurate growth curve of the mass distribution, which of
course should increase beyond one arc-minute.  }
\end{center}
\end{table*}
The table shows that this mass measurement is uncertain and strongly 
depends on the zero-point. Unfortunately, an accurate calibration is not
possible, unless a complete $J$ and $K$ photometric coverage of the total
  FORS field provides the photometric redshifts of the galaxies with $z>0.4$
located far beyond the cluster centre.  Therefore, we cannot
expect better results from magnification bias
  than the gravitational shear ones until then.
\\
The mass from depletion has many sources of errors: Poisson noise
statistics on very low numbers and galaxy clustering which seems obvious 
from the bump observe between $200<r<400$ pixels, whose a fraction comes
from a background lensed cluster at redshift $\approx 1$. Unfortunately,
 the zero-point looks a critical issue. The lower limit  of the density
 of foregrounds seems rather well determined. The intersection of the
depletion of the FORS field is a good indication since this curve should
have an inflexion because it cannot decrease down to zero. This means
that the value of the foreground has probably been minimized. If so,
then the density of galaxies with redshift larger than 0.4 could be
lowered.\\
Alternatively, the position of the background galaxy could be estimated
by simply offsetting the depletion curve of $z>0.4$ galaxies in order to
superimpose to the FORS curve. A crude estimate of the offset, assuming
$Frg$ is settled to 1 $10^{-4}$, leads to $k=2.$. As shown in table 2, 
this strongly lower the total mass. 

\section{Discussion}

It is interesting to  compare our results with X-ray and virial analyses. From 
Fig. 4 of Lewis et al, we see that the mass inside one arc-minute inferred from 
the X-ray emissivity is $M_X(175h^{-1}kpc) \approx 6. \times 10^{13} h^{-1} 
M_{\odot} $,  that is between 1.8 to 2.4 times  lower than the shear analysis, 
$M_{shear}(175h^{-1}kpc)= 1.24\pm 0.17  \ \times  10^{14} h^{-1} M_{\odot} $.
The depletion provides a marginally similar value if $k$ is lower than 2.4, 
but is difficult to reconcile with the X-ray. The depletion approach is, 
however, not sufficiently reliable, since we cannot accurately calibrate 
the zero-point of the photometric redshift sample. It is necessary to
have near infrared data over the whole of the FORS field for this purpose.

On larger scale, the agreement is better. $M_X$ and $M_{shear}$ monotonically 
increase and reach $M_X=1.82^{+0.34}_{-0.23} \times  10^{14} h^{-1} M_{\odot}$
and $M_{shear}=2.3\pm 0.5 \ \times  10^{14} h^{-1} M_{\odot}$ at $r=360 
h^{-1}kpc$. At that radius, which is the limiting distance to which the X-ray
data are reliable, the relative discrepancy is about 20\% is within the errors.
However, even if we assume that the 20\% difference is constant beyond 
$r=360 h^{-1}kpc$, the baryon fraction is not changed significantly with 
respect to the value quoted by Lewis et al (1999). It only decreases from
$f_b=0.18$ to $f_b=0.14$.

We have compared the mass profile inferred from the shear analysis to three 
models (see Fig. \ref{radialprofile}) : a singular isothermal sphere (SIS), an 
isothermal sphere with core radius and the ``universal profile'' (Navarro, 
Frenk \& White 1995). A SIS with velocity dispersion of 
$\sigma=900$ km.sec$^{-1}$ fits the data marginally. The agreement is good at 
large distance, but the discrepancy is important close to the center. A better 
fit is obtained with an isothermal sphere with a core radius. Using the 
parameters of the lens configuration (namely the redshift of the lens and the 
photometric redshift of the lensed galaxies) , the isothermal model with 
 core radius   can be expressed as follows :
\begin{equation}
M(x)=1.28 \times 10^{14} h^{-1} {\rm M}_{\odot} 
\left({\sigma_{\infty} \over 1000}\right)^2 \left({r_c
\over 1'}\right) {x^2 \over \sqrt{1+x^2}}  \ ,
\end{equation}
where $r_c$ is expressed in arc-minutes, $x=r/r_c$, $M(x)$ the mass inside the 
radius $x$ and $\sigma_{\infty}$ the three-dimension velocity dispersion at 
infinity. The best fit predicts  $\sigma=900$ km.sec$^{-1}$  (as expected from 
the fit of a SIS) and  $r_c=45h^{-1}$ kpc. This small value for $r_c$ is robust 
because it is constrained by the change of curvature of the mass profile at 
small radial distance which imposes an upper limit regardless of the mass at 
large radius. The presence of strong lensing features also indicates that matter
is strongly concentrated at the centre and is consistent with the small core
radius. The velocity dispersion is similar within the errors to the galaxy 
velocity dispersion obtained by Carlberg et al (1996) who measured 
1054 km.sec$^{-1}$. This confirms that the method they used to measure the 
velocity dispersion, though it leads to somewhat lower values than previous 
works, is valid and gives a reliable estimator of the dynamical mass.

The universal profile (NFW) fits the mass profile equally well. The NFW profile
may be expressed for this cluster as : 
\begin{equation}
M(x)=1. \times 10^{10} h^{-1} {\rm M}_{\odot} \left({r_s \over 1'}\right)^3
 \delta_c \ m(x)
\end{equation}
where $x=r/r_s$,  $m(x)$ is the dimensionless mass profile (Bartelmann 1996) 
and $\delta_c=\rho_s/\rho_c$, where $\rho_c$ is the critical density. Though 
both the isothermal sphere with core radius and the NFW profiles are within the 
error bars, the latter fits the data better. In particular on small and 
intermediate scales,  the shape  of the radial mass distribution is well 
reproduced by the NFW. The best fit gives $r_s=175h^{-1}$ kpc and 
$\delta_c=3.62 \times 10^4$.

\begin{figure}
\begin{center}
\psfig{figure=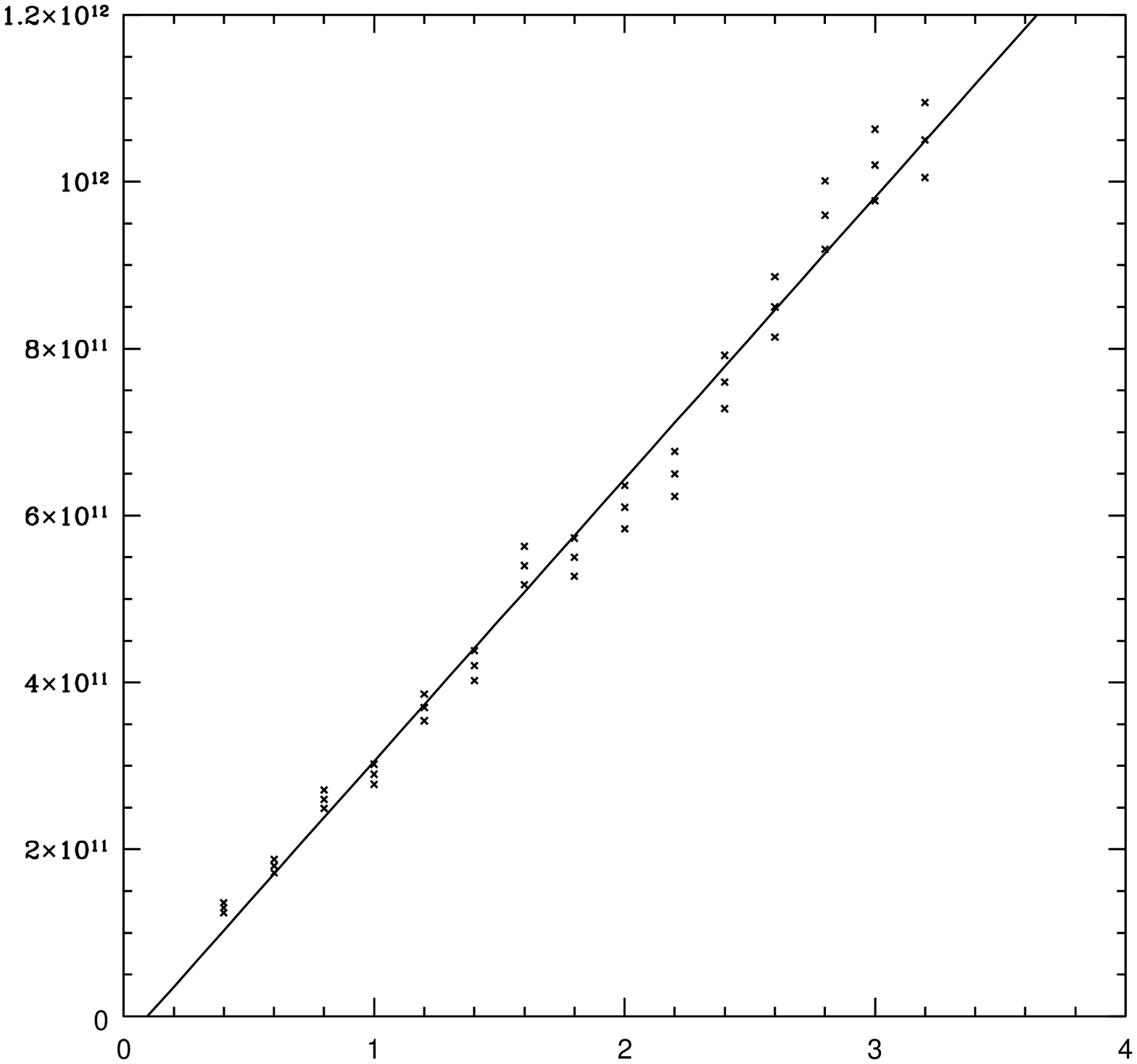,width=8.cm}
\psfig{figure=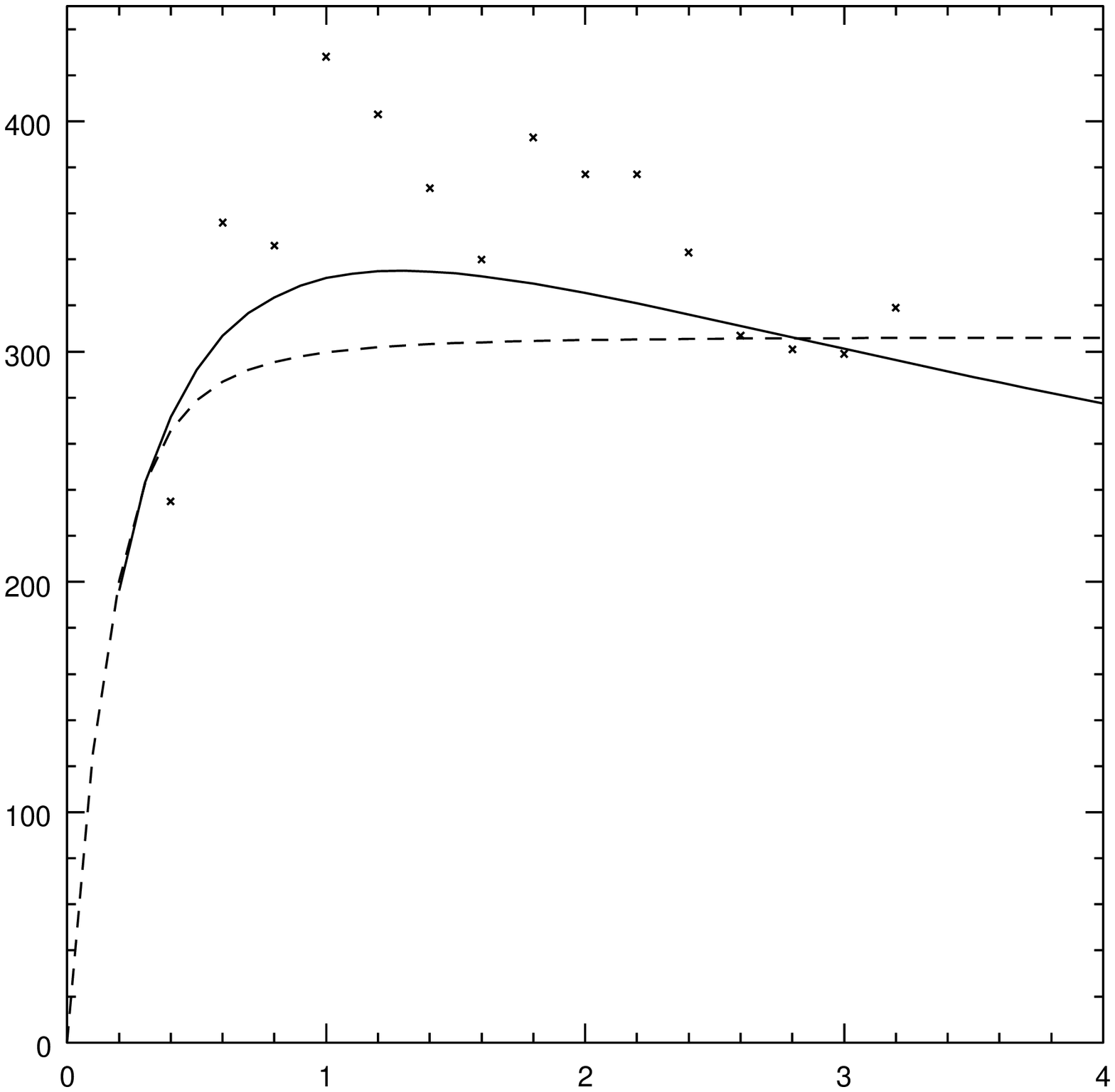,width=8.cm}
\caption{\label{lightrelated}Top panel:Radial distribution of the 
luminosity of cluster galaxies located on the cluster sequence on the 
color-magnitude plot. The profile is remarkably linear.  Bottom panel: radial 
profile of the mass-to-light ratio of  MS 1008-1224 inferred from weak lensing 
and the luminosity of galaxies. The lines represent the best fit Isothermal 
Sphere with a core radius (dashed) and NFW (solid) profiles. 
}
\end{center}
\end{figure}

Fig. \ref{lightrelated} shows the radial luminosity profile of the cluster
galaxies (selected from the colour-magnitude plot). The luminosity was computed 
assuming non-evolution models and by using a $K$-correction of elliptical/S0s 
galaxies for the whole sample. This is a reasonable approximation since the 
galaxies selected with the color-magnitude diagram are mainly early-type 
systems and contains the brightest galaxies which contribute most of the 
luminosity of the cluster.  The error bars in Fig. \ref{lightrelated} assumed 
a constant magnitude error of $\delta_I =0.1$ regardless the signal-to-noise of 
the individual galaxies. This conservative estimate makes allowance for the
underestimation of the $K$correction for late-type galaxies of the sample. The 
$K$-corrections have been computed from the most recent Bruzual \& Charlot 
models (Bruzual \& Charlot 1993).

The light profile is remarkably linear. Hoesktra et al (1998) found similar 
results in Cl1358+62. The best fit to the profile gives a slope $a=3.38\pm0.09 
\times 10^{11} \ h^{-1} \ L_{{\odot}R} /arcmin$ and $b=-0.032\pm0.029 \times 
10^{12} \ h^{-2} L_{{\odot}R}$ (that is, almost zero, as expected). Assuming 
that the mass-to-light,$M/L$ ratio is constant over the field, this linear 
profile  indicates that an isothermal sphere is an acceptable model for the
lens at least up to a radial distance of 700 $h^{-1}$ kpc. However, the profile 
of the mass-to-light ratio, $M/L(r)$, is more puzzling and does not bear this 
out.

The integrated $M/L$ over the field is $M/L =319h$, in excellent agreement with 
the value of the CNOC analysis ($M/L=312h$, Carlberg et al 1996). Fig. 
\ref{lightrelated} shows its radial profile. The lines are the $M/L$ profiles 
computed from the best fit of the mass distributions (isothermal models or NFW) 
divided by  the best fit of the light distribution (straight line). Clearly,
the mass-to-light ratio depends on the radial distance. In the center, it is 
lower than the average, which  expresses the dominating contribution of the cD 
galaxy to the total luminosity and the fact that this cD is much brighter than 
normal ellipticals. At large distance, the mass-to-light ratio seems to converge
towards a constant value. 

The overall shape of $M/L(r)$  is better reproduced by the NFW profile, 
basically because, if the light distribution increases linearly with radius, the
$M/L$ of the NFW profile varies as ${\rm Log}(x)/x$ which shows a maximum at
intermediate scale. Therefore, if the $M/L$ is really not constant with radius 
it would favor the NFW profile against the isothermal sphere. Unfortunately, 
the error bars are large enough to be consistent with both the profiles.
It is worth noting that both profiles underestimate the $M/L$ on intermediate 
scales.  The origin of the  discrepancy  could be the clustering of sources. In
particular, the  second cluster at redshift 0.9 certainly increases the 
amplification and the gravitational shear of galaxies having a redshift larger 
than one and which are located inside the radius $r<300$ pixels (one 
arc-minute). Thus, all the galaxies beyond $z=0.9$ are magnified twice. From 
the depletion point of view, the most distant galaxies are deflected twice, 
which increases the depth and the angular size of the depleted area. From the 
gravitational shear point of view, the distant cluster also enhances the 
distortion that  the weak lensing analysis mistakenly interprets as a strong 
gravitational effect of MS 1008-1224. This could explain why the mass from the 
weak lensing analysis, and therefore the radial distribution of the 
mass-to-light ratio shown in Fig. \ref{lightrelated}, increase rapidly at small 
radius ($r < 300$ pixels) despite a linear increase of the cluster luminosity. 
A similar effect is also discernable in the depletion which has a very 
steep growth curve. 

The discrepancy between X-ray and lensing mass only appears on small scales. But
it is a factor of 2, which is significantly lower than the factor 3.7 obtained 
by Wu \& Fang (1997) from the analysis of strong lensing features. The decrease 
of the discrepancy with radius seems to be a general trend which has already 
been reported (Athreya et al 1999, Lewis 1999, see Mellier 1999 and references 
therein). It must be noted that in most of the studies the comparison has been
done between X-ray and strong lensing features and not weak lensing analysis.

Some of the discrepancy observed in MS 1008-1224 can be produced by the 
distant cluster we have discovered behind it by biasing the mass estimate for 
MS 1008-1224 towards a higher value. However, such a projection effect, similar 
to those discussed by Reblinsky and Bartelmann (1999),  cannot explain a 
factor of 2 discrepancy because the distant cluster is only a tiny fraction of 
the lensed area. It is confined to just one quadrant of the ISAAC field, where, 
additionally, only  background galaxies with redshift higher than 0.9 are 
magnified twice.  The magnitude of its impact on the mass estimate is roughly 
the ratio 
\begin{equation}
{1 \over 4} \ {\left<{D_{ls} \over D_{os}}\right>^{-1}_{z_l=0.9} 
\over
\left<{D_{ls} \over D_{os}}\right>^{-1}_{z_l=0.3}}
 \times {n_{z>0.9} \over n_{z>0.3}} \ , 
\end{equation}
where $n_{z>z_l}$ is the fraction of lensed galaxies with redshift larger than 
$z_l$ and the factor 1/4 is the fraction of the ISAAC area covered by the 
cluster. Using photometric redshifts, we estimate that this ratio is about 30
per cent, as predicted by Reblinsky and Bartelmann (1999). In fact, as it is
obvious from the mass reconstructions, though we see an extension westward 
of the mass map at the position of the cluster, no prominent clump of matter 
is visible and the distortion of the mass maps generated by this 
second lens seems rather weak.

It is worth noting that apart from this distant cluster, contaminations
by other projection effects are not visible at the center, where photometric 
redshifts provide a good idea of the clustering along the
line of sight. The total area covered by ISAAC encompass the region 
where strong lensing features are visible,  where the mass estimate 
from lensing exceeds the X-ray prediction. We find no evidence that biases like
the ones proposed by Cen (1997) or Metzler (1999) are significant in the
central region. In contrast, there is some evidence that the innermost regions 
of MS 1008-1224 is complex, which makes the modeling of the hot gas a 
difficult task. In particular, there is compelling evidence that the center of 
mass does not coincide exactly with the cD galxy :
\begin{itemize}
\item Lewis et al (1999) reported an offset of the X-ray centroid about 15 
arcseconds north of the cD galaxy.
\item the mass reconstructions show an offset of 15 arcseconds north and 15 
arcseconds west for the low resolution mass maps,
\item the depletion area is offseted by 15 arcseconds north and 10
arcseconds west with respect to the cD.
\end{itemize}
While each of the above results are not highly significant in themselves, the
fact that all of them independently point to the offset, and in fact in the
same direction strongly suggests that the cD is not at the centroid of the
mass distribution. As seen in our high resolution mass reconstruction, it may
coincide with the lower subclump.
The isoluminosity and number density contours clearly show that the 
light distribution is clumpy, in particular along the north direction,
as in the mass maps.
If those clumps overlap along the line of sight they will look like merging 
system by projection effect and the projected mass density produced by
the weak lensing analysis  will be centered on the projected centre 
of mass, if the resolution is too low to separate each component. In fact, 
the X-ray, the luminosity and the number density contours show elongated and 
clumpy filaments along the north direction which all indicate that MS
1008-1224 is still experiencing merging processes.  If this assumption
is valid, then the hot gas is not in equilibrium. The merging process
produces shocks and gas flows between clumps, such as those seen in Schindler 
\& M\"uller' simulations (1993) or those reported by  Kneib et al (1996) and 
Neumann \& B\"ohringer (1999) in the lensing cluster A2218.  Athreya et al 
(1999) reported similar trends in A370, with remarkable similarity with MS 
1008-1224 : a good agreement with X-ray and weak lensing analysis on large scale
and a discrepancy of a factor of 2 also in the inner region. Since A370 is 
clearly composed of merging clumps, Athreya et al also interpret the 
discrepancy as  a result of an oversimplification of the physics of the hot gas.
We then suspect that the X-ray analysis oversimplified the dynamical stage of 
the gas inside MS 1008-1224, producing a wrong estimate of the mass and thus
a discrepancy between the X-ray and the weak/strong lensing analysis. This, as 
suggested earlier by Miralda-Escud\'e and Babul (1995), easily explains the good
agreement on large scale between the weak lensing, the X-ray and also the 
virial mass (see Lewis et al 1999) and the apparent contradiction between X-ray 
and strong lensing.

\section{Conclusion}
Thanks to the deep multicolor subarcsecond images obtained with 
FORS and ISAAC by the Science Verification Team, it has been 
possible to   map the mass distribution of the lensing cluster
MS 1008-1224, to evaluate the reliability of non-parametric
reconstruction and to scale the mass carefully.  
The comparison between the mass map from weak lensing analysis,
the X-ray reconstruction and the light distribution from optical 
data lead to several interesting insights.

\begin{itemize}
\item the weak lensing analysis of the FORS data provide stable and 
reliable mass maps which look similar in B, V, R and I filters. This
shows that the PSF correction and the mass reconstruction algorithms 
work  well.

\item good BVRIJK photometry allowed for photometric redshift estimation which 
in turn allowed us to obtain the absolute mass from weak-lensing.

\item using the sample with photometric redshift, we discovered a very distant
cluster of galaxies located behind MS 1008-1224. This is the first 
observational evidence of cluster-cluster lensing.

\item unfortunately, this lensed cluster partly  compromises the use of the
depletion curve to estimate the mass of the cluster independent of the weak 
shear analysis.  The slope of the growth curve shows irregularities produced 
by the excess of galaxies in the lensed clusters. This kind of clustering is
is an intrinsic limitation of the practical usefulness of the magnification 
bias which has already been stressed by Fort et al (1996) and Hoesktra et al
(1999).

\item On a more optimistic view,  MS1008-1224 can be used as a gravitational 
telescope in order to study the population of a cluster of galaxies at redshift
one. Thanks to the magnification, the number density contrast of the distant 
cluster increases and many more galaxies than the usual fraction visible in 
such distant clusters can be studied. We have not dwelt on this point since it 
is far beyond the scope of this work; but a join multicolor and spectroscopic 
analysis of this cluster could be valuable.

\item On large scales, the total mass inferred from weak lensing is in good 
agreement with the X-ray analysis. In contrast, we still have a considerable
discrepancy on small scales.  The clumpiness of the light distribution, the 
elongated shape, the X-ray emissivity and the double peak mass contours are 
indications that the cluster is still in the throes of a merger.  We therefore 
believe that the X-ray gas is not in equilibrium in the innermost part of the 
cluster and that this is the principal reason for the mass discrepancy.
\end{itemize}

In order to go further in the analysis of this cluster we need first a larger 
coverage of the FORS field in the infrared. This will provide us photometric 
redshifts of galaxies beyond one arcminute from the cluster center and hence a
better estimate of the zero point of the depletion curve.
It would be also valuable to get HST images of the center in order to model the
innermost regions using strong lensing features. This will provide with a better
accuracy the cluster center and can be used to check if the contradiction 
between the X-ray and the lensing mass can be sorted out by a more accurate 
lensing model of the cluster center. Photometric redshifts will be particularly 
helpful since they will provide the redshift of each arc(let). Finally, it 
would be interesting to get the spectrum of the arc candidate \#4. If confirmed 
as a gravitational arc, then its position and curvature would immediately imply
that the center of mass of the cluster is not located on the cD galaxy.

We believe that such a detailed weak-lensing analysis should be carried out on
many clusters of galaxies  with ground-based optical telescopes. However, the
technical requirements for such an investigation are stringent. One must 
combine very deep observations, subarcsecond  images and multicolor photometry,
from the B-band up to the infrared K-band. The exceptional combination of FORS 
and ISAAC on the VLT is the best tool available at present for such a project.
It is only after doing similar investigations on a large sample of cluster, 
that we may be able to obtain a clear understanding of the amount and 
distribution  of dark matter in clusters, understand the reasons behind the
the X-ray/lensing mass discrepancy and in general to use weak lensing analysis 
of clusters as a reliable cosmological tool.

{
\acknowledgements  We thank H. Hoesktra for providing his own 
 updated version of the IMCAT software and
for the numerous discussions we had together on weak lensing.  We thank
 also E. Bertin, T. Erben, N. Kaiser, R. Maoli, D. Pogosyan and P. Schneider 
   for fruitful discussions on lensing and data analysis.  We would like
to thank also the Science Verification Team of FORS and ISAAC at ESO for
the remarkable work they did in order to provide these data to the 
 ESO community.\\
 The TERAPIX data center  provided its computing facilities for 
  the data reduction, the  lensing analyses and the simulations. 
  This work was supported the  TMR network ``Gravitational Lensing:
 New Constraints on Cosmology'' and the Distribution of Dark Matter'' 
 of the EC under contract No. ERBFMRX-CT97-0172 and the Indo-French 
 Centre for the Promotion of Advanced Research IFCPAR grant 1410-2.
 }


\end{document}